\documentclass[twocolumn,trackchanges]{aastex61}
\usepackage{mathptmx}
\usepackage{epstopdf}
\usepackage{color}
\usepackage{float}
\usepackage[hyphenbreaks]{breakurl}
\usepackage{graphicx}
\usepackage{subfigure}
\usepackage{tabularx} 
\usepackage{color}

\newcommand{\msun}{\rm M_{\sun}}

\def \micron {$\mu$m}

\newcommand{\ergs}{\hbox{erg s$^{-1}$}}              
\newcommand{\kms}{\hbox{km s$^{-1}$}}          
\newcommand{\halpha}{\rm H{$_\alpha$}}
\newcommand{\hbeta}{\rm H{$_\beta$}}
\newcommand{\oiii}{[OIII]}
\newcommand{\oii}{[OII]}

\newcommand{\nii}{[NII]}
\newcommand{\sii}{[SII]}
\newcommand{\dproj}{D$_{\rm proj}$}
\newcommand{\voff}{$\Delta_{\rm V}$}

\usepackage{CJK}

\begin{document}
\begin{CJK}{GB}{gbsn}

\bibliographystyle{apj}

\shorttitle{Emission Line Galaxy Pairs}
\shortauthors{Dai et al.}
\received{2021, Jan 11}
\revised{2021, Sep 19}
\accepted{2021, Oct 11}

\title{Spectroscopically Identified Emission Line Galaxy Pairs in the WISP survey\footnote{Based on observations 
made with the NASA/ESA Hubble Space Telescope,
which is operated by the Association of Universities for Research in
Astronomy, Inc., under NASA contract NAS 5-26555. 
These observations are associated with programs 
11696, 12283, 12568, 12092, 13352, 13517, and 14178.} }

\author[0000-0002-7928-416X]{Y.Sophia Dai
(´÷êÅ)}
\affil{Chinese Academy of Sciences South America Center for Astronomy (CASSACA), 
National Astronomical Observatories of China (NAOC), 20A Datun Road, Beijing, 100012, China}
\footnote{Corresponding author: daysophia$@$gmail.com}

\author{Matthew M. Malkan}
\affil{UCLA Department of Physics and Astronomy, Los Angeles, CA 90095-1547, USA}

\author{Harry I. Teplitz}
\affil{IPAC, Mail Code 314-6, Caltech, 1200 E. California Blvd., Pasadena, CA 91125, USA}

\author{Claudia Scarlata}
\affil{Institute for Astrophysics, University of Minnesota, 116 Church St. SE, Minneapolis, MN 55455, USA}

\author{Anahita Alavi}
\affil{IPAC, California Institute of Technology, 1200 E. California Boulevard, Pasadena, CA 91125, USA}

\author{Hakim Atek}
\affil{Sorbonne Universit\'e, CNRS UMR 7095, Institut d'Astrophysique de Paris, 98bis bvd Arago, 75014, Paris, France}

\author{Micaela Bagley}
\affil{College of Natural Sciences, The University of Texas at Austin, 2515 Speedway, Austin, TX 78712, USA}

\author{Ivano Baronchelli}
\affil{Dipartimento di Fisica e Astronomia, Universit`a di Padova, vicolo Osservatorio, 3, 35122 Padova, Italy}

\author{Andrew Battisti}
\affil{Research School of Astronomy and Astrophysics, Australian National University, Cotter Road, Weston Creek, ACT 2611, Australia}

\author{Andrew J Bunker}
\affil{Sub-department of Astrophysics, Department of Physics, University of Oxford, Denys Wilkinson Building, Keble Road, Oxford OX1 3RH, UK}
\affil{Kavli Institute for the Physics and Mathematics of the Universe (WPI), The University of Tokyo Institutes for Advanced Study, The University of Tokyo, Kashiwa, Chiba 277-8583, Japan}

\author{Nimish P. Hathi}
\author{Alaina Henry}
\affil{Space Telescope Science Institute, 3700 San Martin Drive, Baltimore, MD, 21218, USA}

\author{Jiasheng Huang}
\affil{Chinese Academy of Sciences South America Center for Astronomy (CASSACA), 
National Astronomical Observatories of China (NAOC), 20A Datun Road, Beijing, 100012, China}

\author{Gaoxiang Jin}
\affil{Chinese Academy of Sciences South America Center for Astronomy (CASSACA), 
National Astronomical Observatories of China (NAOC), 20A Datun Road, Beijing, 100012, China}

\author{Zijian Li}
\affil{Chinese Academy of Sciences South America Center for Astronomy (CASSACA), 
National Astronomical Observatories of China (NAOC), 20A Datun Road, Beijing, 100012, China}

\author{Crystal Martin}
\affil{Department of Physics, Broida Hall, University of California Santa Barbara, CA 93106, USA}

\author{Vihang Mehta}
\author{John Phillips}
\affil{Institute for Astrophysics, University of Minnesota, 116 Church St. SE, Minneapolis, MN 55455, USA}

\author{Marc Rafelski}
\affil{Space Telescope Science Institute, 3700 San Martin Drive, Baltimore, MD, 21218, USA}
\affil{Department of Physics and Astronomy, Johns Hopkins University, Baltimore, MD 21218, USA}

\author{Michael Rutkowski}
\affil{Institute for Astrophysics, University of Minnesota, 116 Church St. SE, Minneapolis, MN 55455, USA}

\author{Hai Xu}
\author{Cong K Xu}
\affil{Chinese Academy of Sciences South America Center for Astronomy (CASSACA), 
National Astronomical Observatories of China (NAOC), 20A Datun Road, Beijing, 100012, China}

\author{Anita Zanella}
\affil{Istituto Nazionale di Astrofisica, Vicolo dell'Osservatorio 5, 35122 Padova, Italy}

\begin{abstract}
We identify a sample of spectroscopically measured emission line galaxy (ELG) pairs up
to $z=$1.6 from the WFC3 Infrared Spectroscopic Parallels (WISP) survey. 
WISP obtained slitless, near-infrared grism spectroscopy 
along with direct imaging in the J and H bands by observing in the pure-parallel mode with the Wide
Field Camera Three (WFC3) on the Hubble Space Telescope ($HST$). 
From our search of 419 WISP fields covering an area of $\sim\,0.5\,{\rm deg}^{2}$, 
we find 413 ELG pair systems, mostly \halpha\ emitters. 
We then derive reliable star formation rates (SFRs) based on 
the attenuation-corrected \halpha\ fluxes. 
Compared to isolated galaxies, we find an average SFR enhancement of 40\%-65\%,
which is stronger for major pairs and pairs with smaller velocity separations ($\Delta_v < 300\,\kms$). 
Based on the stacked spectra from various subsamples, 
we study the trends of emission line ratios in pairs, 
and find a general consistency with enhanced lower-ionization lines. 
We study the pair fraction among ELGs,
and find a marginally significant increase with redshift $f \propto\,(1+z)^\alpha$,
where the power-law index $\alpha\,=$\,0.58$\pm$0.17 from $z\sim0.2$ to $z\sim$1.6.
The fraction of Active galactic Nuclei (AGNs), 
is found to be the same in the ELG pairs 
as compared to isolated ELGs. 
\end{abstract}

\section{INTRODUCTION}
Mergers and interactions play 
a key role in galaxy evolution.
In addition to large scale accretion of baryonic and dark matter \citep[e.g.][]{dimatteo08, dekel09, steidel10, bournaud11},
galaxy mergers can convert gas into stars,
and feed the growing of supermassive black holes
\citep[e.g.][]{whitenrees78, mihosnhernquist96, snm96}. 
Both simulations and observations 
suggest that 
galaxy interactions elevate the star formation,
especially in the center of the galaxy \citep[e.g.][]{snm96, dimatteo07, ellison08, fensch17}.
The degree of star formation rate (SFR) enhancement may depend 
on galaxy mass ratios, separation of the galaxies, 
and gas amount \citep[e.g.][]{cox08, patton13, scudder15, davies15, moreno20}. 
Locally, the SFR enhancement has been confirmed 
in large statistical samples, 
with a pair separation up to 150\, kpc \citep{patton13, ellison13, violino18}, 
when compared to a control sample of isolated galaxies with similar stellar mass. 
At higher redshifts, however, the situation is less clear due to limited 
observations and identifications of pair samples,
though controversies exist as to whether mergers
are the main driver of star formation and mass assembly 
since $z\sim$4 \citep[e.g.][]{deravel09, williams11, wuyts11, tasca14}.  
 
Given its fundamental importance
to galaxy assembly and size evolution over cosmic time,
many investigators have tried to measure the galaxy merger rate.
Due to the large uncertainties associated with
the merging timescale, 
wide pairs are often used,
assuming they will merge at some point in the future.
Observational and theoretical studies have shown that
pair fractions, and thus merger rates, 
depend on mass ratios, 
luminosities, 
and optical colors 
 \citep[e.g.][]{pattonnatfield08, hopkins10, lotz11, keenan14, lopezsanjuan15}.
For instance, the major merger rate appears to evolve as $(1+z)^{\alpha}$,
where $\alpha=$2-3,  
as predicted by simulations and confirmed by observations \citep[e.g.][]{bridge10, xu12a, rodriguezgomez15, 
lopezsanjuan15, man16, ventou17, duncan19},
at least up to a possible peak around $z\sim$3 
\citep[e.g.][]{conselicenarnold09, ventou17, qu17, kaviraj15, mantha18}.
Others have found flatter or close-to-constant merger fraction,  
especially among massive galaxies ($logM_* > 10.3\,\msun$), 
as in \citet[][$\alpha = {0.8\pm0.2}$, $z =0.5-6$]{duncan19}. 
and in \citet[][$\alpha = {-0.4\pm0.6}$, $z=0.4-2$]{williams11}. 
These observational differences can be 
attributed to  different selection effects and 
the different conversion factors between pair fraction and merger rate \citep{mantha18, duncan19}. 
For instance, after converting the observed pair fraction to merger rate,
a constant merger rate was found up to z$\sim$3 in simulations \citep{snyder17},
and to $z\sim$6 in observations \citep{duncan19}.
At higher redshifts ($z>$3), a steady decrease is noticed 
by both simulations up to z$\sim$4 \citep[e.g.][]{snyder17} 
and observations up to $z\sim\,6$ \citep[e.g.][]{ventou17}. 
In general, up to $z\sim2.5$, pair fractions 
are found to be 2-16\% for major mergers (mass ratio $<$ 4),
and $\sim20-30\%$ if minor mergers (mass ratio $>$ 4) are included 
 \citep[e.g.][]{ellison08, keenan14, man16, zanella19, mantha18}.

 The uncertainties in the merger rate measurement at $z\ge1$ 
are caused by various factors, 
including sample selection, pair morphology and 
merging timescales \citep[e.g.][]{law15, lopezsanjuan15, snyder17, mantha18, duncan19}.
At $z >$ 0.5, 
pair identification becomes more difficult, 
mostly related to the increasing uncertainties in photometric redshifts
and declining resolution for morphology identifications. 
High spatial resolution imaging 
and accurate redshift information, preferably spectroscopic $z$,
are desired.
False pair identification is high ($>$50\%) 
from morphological identification alone,
due to chance sky alignments \citep[e.g.][]{pattonnatfield08, chou12, law12, law15}.
Ground-based spectroscopy is only available for small samples \citep[e.g.][]{law15},
or limited to local galaxies ($z < $ 0.2), 
as large separations are required for multi-object spectrographs \citep[e.g.][]{ellison08}.
For instance, SDSS pairs are 
biased to large separation ($>$55\arcsec) systems to avoid slit/fiber collisions. 
\citet{xu12a} studied 
the merger rates of close major-merger pairs using a 
K-band selected local ($z\sim$ 0) sample and a sample 
of $0.2 \leq z \leq 1$ pairs selected using high quality photo-z data 
in the COSMOS field, both having high completeness and reliability. 
On the other hand,
the $z>$ 1 studies,  mostly based on estimates for massive and luminous galaxies 
\citep[e.g.][]{bundy09, bridge10, williams11},
usually have large uncertainties due to small sample sizes 
and incomplete spectroscopic redshifts \citep[e.g.][]{bluck12}.
Compared to major mergers selected by stellar mass ratios,
those selected by optical flux ratios show 
a systematically higher and increasing merger fraction \citep{lotz11, man16, mantha18}.

Earlier studies of spectroscopically identified $z>1$ pairs,
though limited, 
have found generally consistent properties with statistical, photometric samples. 
In a study of 113 spectroscopic pairs from the 
deep MUSE (Multi Unit Spectroscopic Explorer) observations in the Hubble Ultra Deep Field (HUDF), 
\citet{ventou17} show that the fraction of pairs increases up to $z\sim$3 and then slowly decreases.
The star formation conditions, however, seem to diverge in different spectroscopic pair samples.
At $z<1$, \citet{wong11} found that tidal
interactions are responsible for a 15-20\% increase of specific SFR (sSFR $=$ SFR/M$_*$) 
in pairs, as compared to isolated galaxies. 
In this study based on the Prism Multi-Object Survey (PRIMUS),
no significant redshift dependence was found. 
At $z>2$, however, a small number (2) of spectroscopically-confirmed pairs show
similar star forming properties as $z\sim$2 main-sequence galaxies \citep{law15}. 
Based on 30 spectroscopic pairs from the MOSFIRE Deep Evolution Field (MOSDEF) Survey \citep{kriek15},
\citet{wilson19} also found no measurable SFR enhancement or metallicity deficit 
for $1.4 < z < 3.8$ pairs as compared to isolated galaxies with similar stellar masses. 
This can be explained by the earlier merger stages (e.g. pre-coalescence) before
the triggering of starburst \citep{bustamante18}. 

To reconcile the uncertainties associated with photometric redshifts
and declining resolution for morphology identifications at $z\,>$0.5,
a statistically significant sample of spectroscopically confirmed
galaxy pairs is needed. 
In this paper, we search for galaxy pairs 
using the high spatial resolution spectra in 
the Hubble Space Telescope's (HST) WFC3 Infrared Spectroscopic Parallel 
survey (WISP, PI: M. Malkan, GO\# 11696, 12283, 12568, 12902, 13352, 13517, 14178) \citep{atek10}, 
which includes $\sim$9,000 high signal-to-noise (S/N) emission line galaxies (ELGs) with spectroscopic redshifts. 
WISP is a Hubble Space Telescope pure parallel survey,
which observes ``random" parallel fields with the Wide Field Camera 3 (WFC3), 
obtaining
direct imaging (IR, and sometimes UVIS) and IR grism spectroscopy (0.8-1.7\,\micron) simultaneously. 
WISP detects emission line galaxies (ELGs) without preselection, allowing
one of the first spectroscopic studies of 
faint, possibly low-mass, sometimes low-metallicity
ELG pairs up to $z=$1.5. 
Many of the WISP-discovered strong ELGs,
with their high equivalent width (EW),
are the local analogs of the sources of re-ionization at $z>$6 \citep{atek14}. 

We introduce the ELG pair sample in Section~\ref{sec:sample},
then study their star formation properties in Section~\ref{sec:sfr}.
In Section~\ref{sec:stack}, we stack the pair spectra in different bins and study their emission line ratios,
followed by analysis of the pair fraction and AGN fraction in Section~\ref{sec:frac},
before a brief summary (Section~\ref{sec:summary}).
Throughout, we assume a $\Lambda$-dominated flat universe, 
with $H_{\rm 0} =$70\, km\,s$^{-1}$ Mpc$^{-1}$, 
$\Omega_\Lambda =0.7$, 
$\Omega_{\rm M} =0.3$.All magnitudes are in AB system.

\section{The ELG Pair Sample}
\label{sec:sample}
\subsection{The WISP survey}
The sample of ELG pairs was selected from the WISP survey.
WISP obtained slitless, near-infrared grism spectroscopy 
along with direct imaging in the J and H bands, by observing in the pure-parallel mode with the Wide
Field Camera Three (WFC3) on the HST over $\sim$1000 orbits\footnote{https://archive.stsci.edu/prepds/wisp/}. 
The spectra are obtained with the G102
($\lambda =$0.8-1.15\micron, R$\sim$210) and G141 grisms (($\lambda =$1.075-1.70\micron, R$\sim$130), 
together with direct imaging in the 
J and JH or H bands (F110W and F140W / F160W, respectively). 
Out of the 483 fields observed, F140W is used in 190 fields and F160W in 289 
fields\footnote{Another four fields were observed in F110W only but not used in this work.}. 
For convenience, hereafter the H band is referred to either F140W or F160W band, 
depending on the actual filter used. 
For some fields,  WFC3/UVIS imaging in F606W F814W bands are also available.
The WFC3 data were reduced through a customized pipeline 
updated from \citet{atek10}.
Each field was examined by at least two reviewers 
via a customized interface for redshift identification.

Each WISP field of view covers a projected area of 2.3$\arcmin\,\times\,2.2 \arcmin$,
corresponding to $\sim$350 kpc$^2$ 
at $z=$1--2.  
The WISP HST/WFC3 imaging data have a 
pixel scale of ${\rm 0.13\arcsec\,pix^{-1}}$.
The spectroscopic redshift was measured from the emission lines detected in
the G102 and the G141 grisms.
A typical ELG flux limit in WISP for an \halpha\ emission line (\halpha$_{\rm raw}$, \nii\ not corrected)
is $\sim\,5\times10^{-17}\ergs$ (5$\sigma$). 
More details regarding 
the line identification are given in \S\ref{sec:control}. 

\subsection{The Pair Sample}
\begin{figure*}
\begin{center}
\includegraphics[scale=0.7]{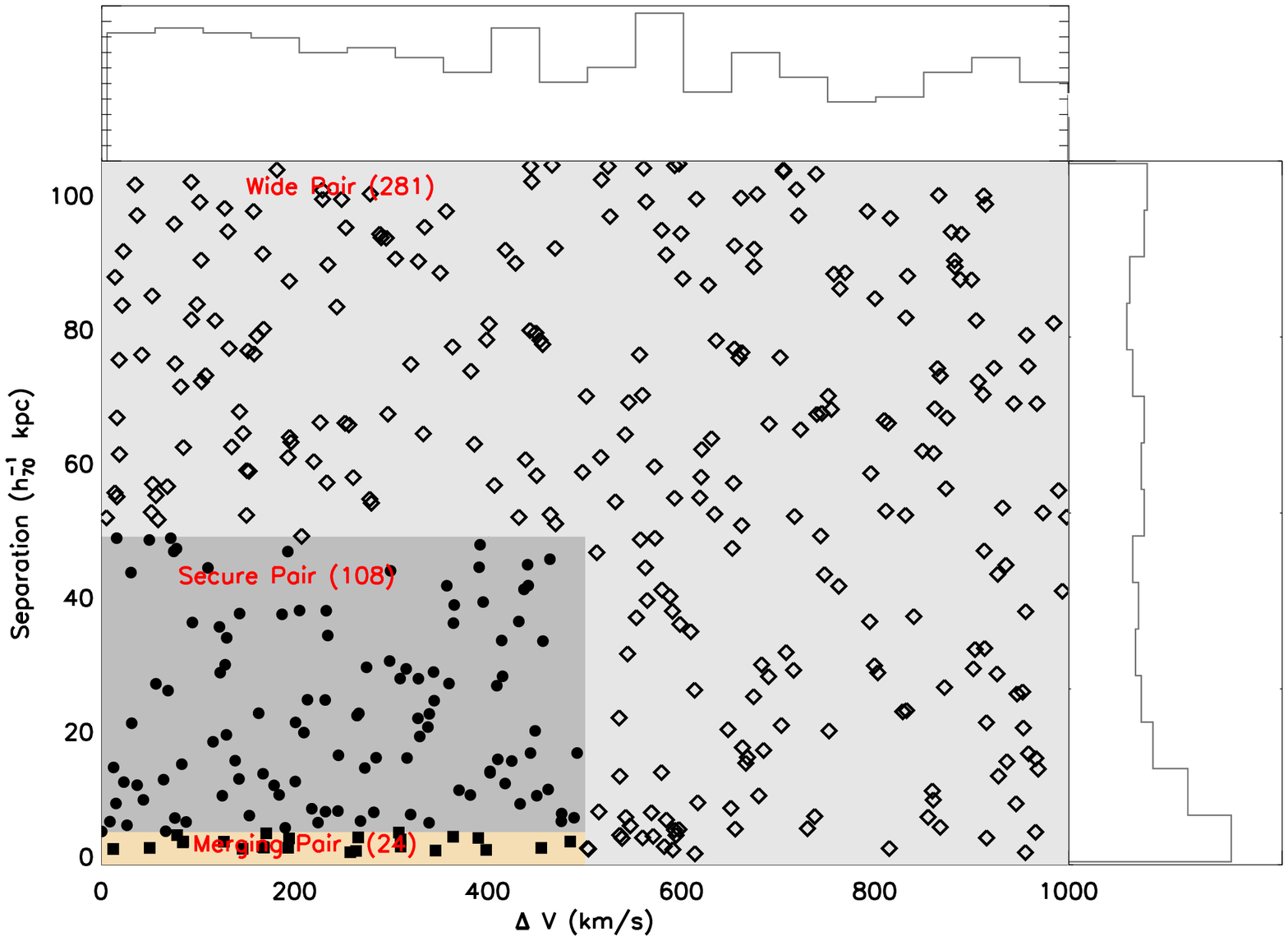} 
\end{center}
\figcaption{Distribution of Emission Line Galaxy pairs 
in the projected separation and relative velocity space.
Histograms in the top and side panels show the distribution in the two axes. 
A total of 413 ELG pair systems are identified,
including 24 merging (filled squares), 108 secure (filled circles), 
and 281 wide pairs (open diamonds; for definition, see \S\ref{sec:sample}). 
Systems with $>$2 members are plotted only once, represented by the closest pairs of the group.
The WFC3 resolution is 0.13\arcsec\,pixel$^{-1}$, while a minimum of 2 pixels are required to 
spatially resolve the galaxies.
At redshift $\sim$1, this corresponds to a resolution of $\sim$1\,kpc.
\label{fig:dist}}
\end{figure*}

We calculated the angular separations based on the sky positions,
and then the projected physical separation \dproj\ (in $h^{-1}\, {\rm kpc}$)
using the average $z$ of the members.
The comoving velocity offset \voff (in $\kms$) was
calculated from the redshifts of the pair members.
All ELG pairs have 
at least an \halpha (6564\AA, \nii\ corrected) or \oiii (5007, 4959\AA) line detected at S/N $>$3.
The pairs are further divided into 3 groups based on their physical separations 
and velocity differences (See also Figure~\ref{fig:dist}):\\
\\
A. Merging Pairs (2 $<$ \dproj $< 5\,h^{-1}\, {\rm kpc}$, \voff $< 500\,\kms$),\\  
B. Secure Pairs (5 $<$ \dproj $ < 50\,h^{-1}\, {\rm kpc}$, \voff $< 500\,\kms$),\\
C. Wide Pairs (\dproj $  < 105\,h^{-1}\, {\rm kpc}$, \voff $< 1000\,\kms$)\footnote{105 $h^{-1}_{70}\, {\rm kpc}$ 
was chosen as an equivalent of 150\,kpc for h$=$ 0.7.
This value is chosen as the separation upper limit 
where SFR enhancement has been previously reported in pairs in the literature
\citep[e.g. in local SDSS galaxy pairs, ][]{patton13}}.
\\
\\
Pairs that are $<$2 pixels apart,
corresponding to a \dproj\ of $\sim$2 kpc at z $=$ 1.5,
are not resolvable with HST's spatial resolution. 
In this pair sample,
the angular separation ranges from 0.28$\arcsec$ to 30.6$\arcsec$,
with a median of 10.3$\arcsec$.
Most of our pair sample consists of compact members,
while 112 systems (27$\pm$3\%) are identified as disturbed systems,
showing evidence of tidal tails or disturbed morphology 
based on visual inspections.

A total of 413 spectroscopically identified ELG pair systems 
are selected from a 
parent sample of 8,192 ELGs in 419 WISP fields. 
The pair sample consists of 24 merging pairs, 108 secure pairs, 
and 281 wide pairs. 
For a fraction (108/413, $\sim$26\%) of the WISP pairs,
Spitzer IRAC observations are also available (PI: Colbert, ID: 80134,
90230,10041,12093). 
For these objects we gathered multi-wavelength data 
from HST (UVIS, J, H), ground based photometry follow-up 
with the Palomar 5.0m, Magellan 6.5m, or WIYN 3.5m telescopes 
( in the u, g, r, i bands),
to the IRAC (3.6\,\micron, 4.5\,\micron) bands (Battisti et al., in prep.; Baronchelli et al., in prep.).
The stellar mass was then estimated using the CIGALE SED-fitting code \citep{noll09, boquien19}, 
assuming a Chabrier IMF, an exponential star formation history in steps of 0.1 dex, 
variable metallicity between 0.004 - 0.05, 
and the \citet{cf00} dust attenuation law.  

Table~\ref{tab:sample1} and Table~\ref{tab:sample2} summarized the basic properties of the pair sample.
The full pairs catalogs are provided in the online version of the paper. 
Figure~\ref{fig:dist} shows the distribution of \dproj\ and \voff\ of the pair sample.
The pair sample does not peak at any specific projected separations or velocity offset,
except for merging pairs which gather at the smallest velocity and physical separations.  

\subsubsection{Major Pair Fraction }
\label{sec:major}
For the subsample of 108 pairs with stellar mass estimates, 
47/108 (44\%) can be classified as major mergers (mass ratio $< $ 4:1) 
and 61/108 (56\%) as minor mergers (mass ratio $>$ 4:1).
In comparison, 
the major merger fraction based on the H band flux ratio for the full sample of 413 pairs
is systematically higher,
at $\sim$ 69\% for major (H band flux ratio $<$  4:1)
and $\sim$ 31\% for minor (H band flux ratio $>$ 4:1) pairs, respectively.
This is consistent with what was found previously in the UltraVISTA/COSMOS pairs, 
in which the fraction of H-band selected major pairs are higher than
stellar mass selected major pairs\citep{man16}. 

\subsubsection{Higher-order Systems}
Among the 413 ELG pairs, 
there are also 47 multiple systems,
including 36 triplets, 
4 quadruple systems, 
and 7 quintuple systems.
This fraction of $\sim 11 \pm 2 \%$ of higher-order systems
is significantly higher than the $\sim$5\% reported in SDSS galaxy pairs \citep{ellison08}.
By design, SDSS and WISP are targeting different samples with different selections,
namely a multi-parameter color selection in SDSS $vs$ WISP's emission line selection. 
In addition, SDSS adopted a  much more complicated pair sample selection (10 criteria)
than in this WISP sample (3 criteria).  
Both can contribute to this significant difference.
Our fraction is otherwise close to the result for ELGs in \citet{zanella19}, 
who found that 13\% of their ELGs have multiple `satellites'.
In the following analyses, 
the multiple systems are counted as one `pair',
similar to the SDSS approach \citep{ellison08}, 
and the properties of the brightest pair members are used
except otherwise noted.

\subsection{The Control sample}
\label{sec:control}
\begin{figure*}
\begin{center}
\includegraphics[scale=0.6]{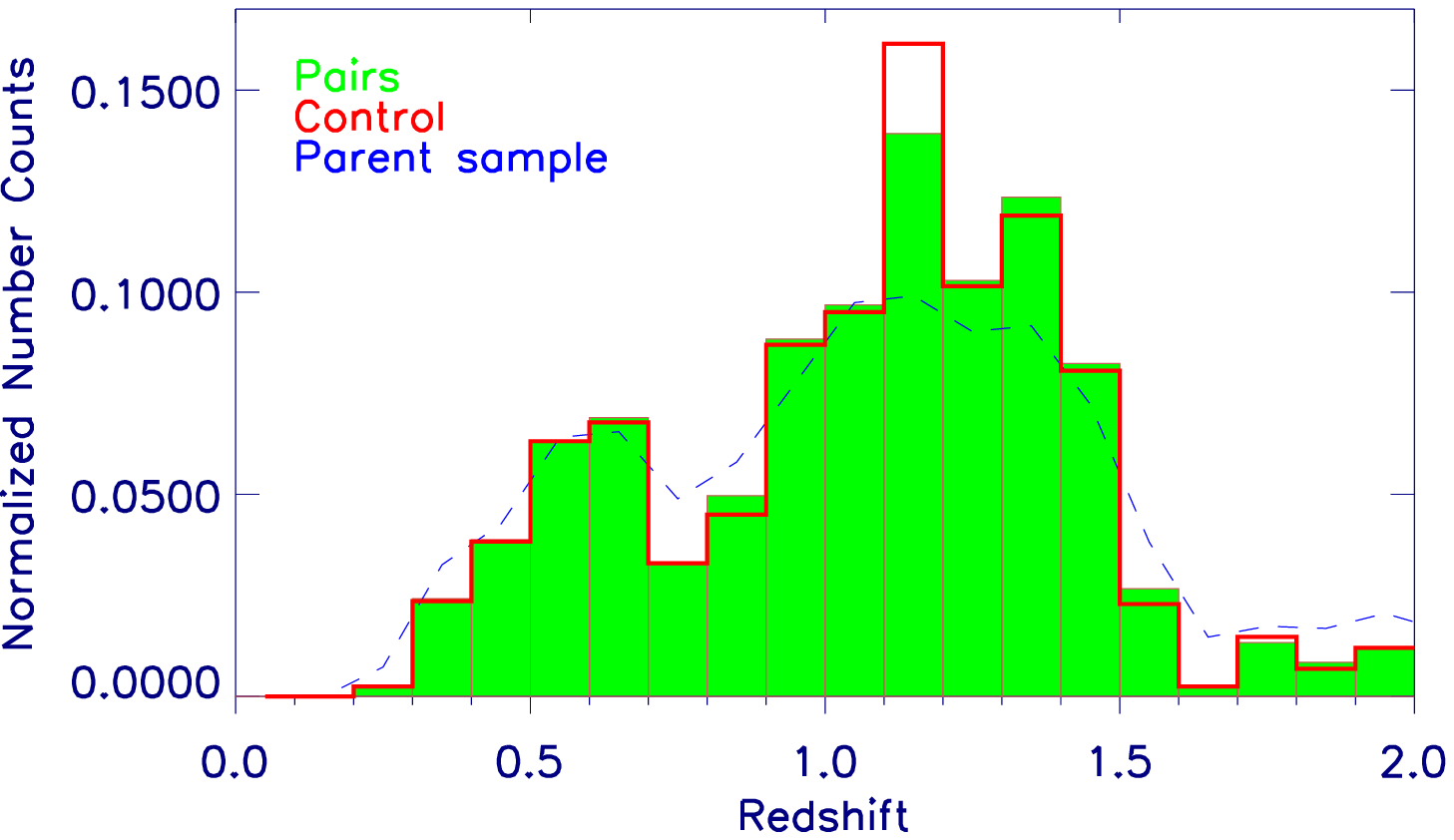} 
\includegraphics[scale=0.6]{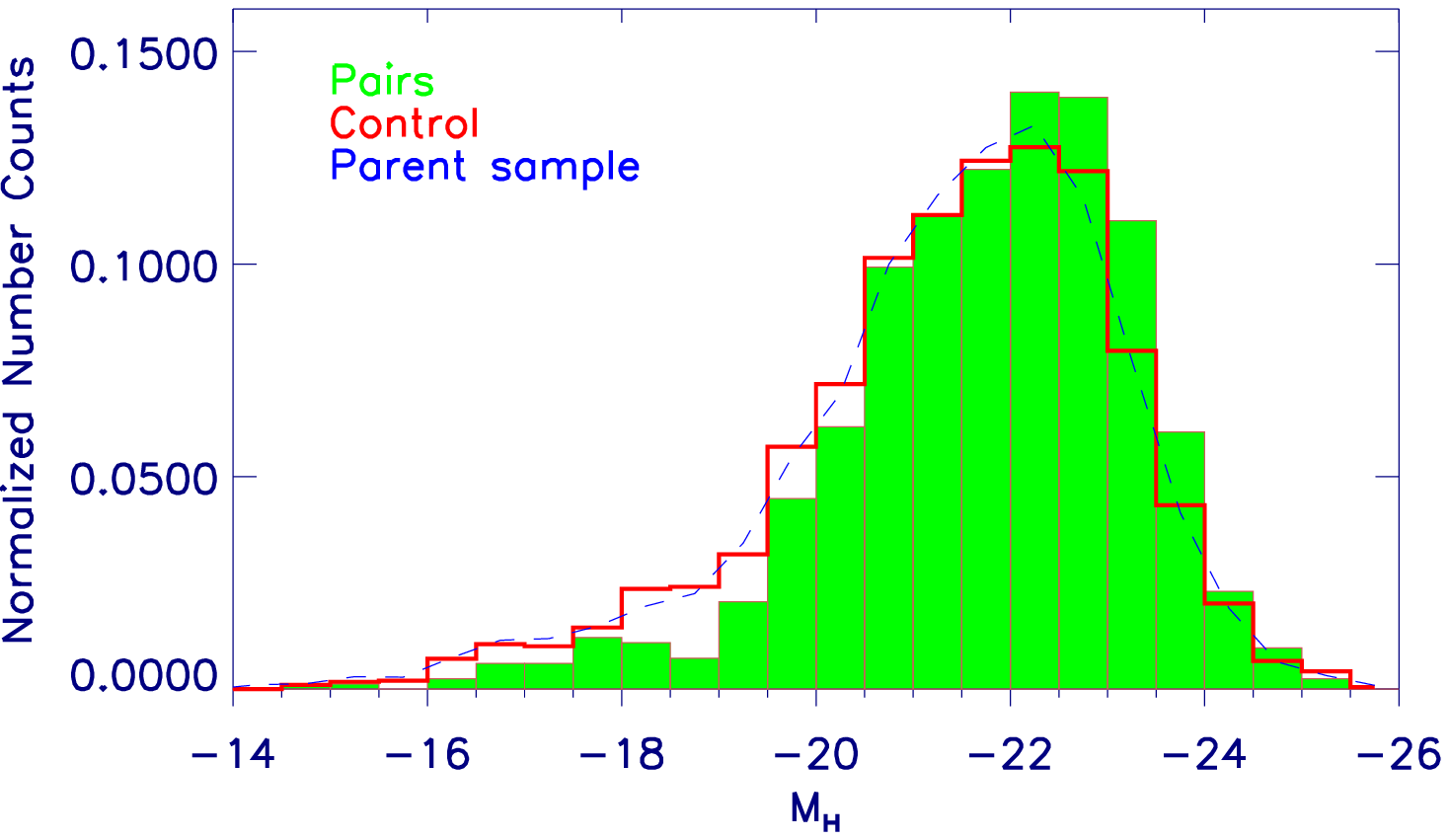} 
\includegraphics[scale=0.6]{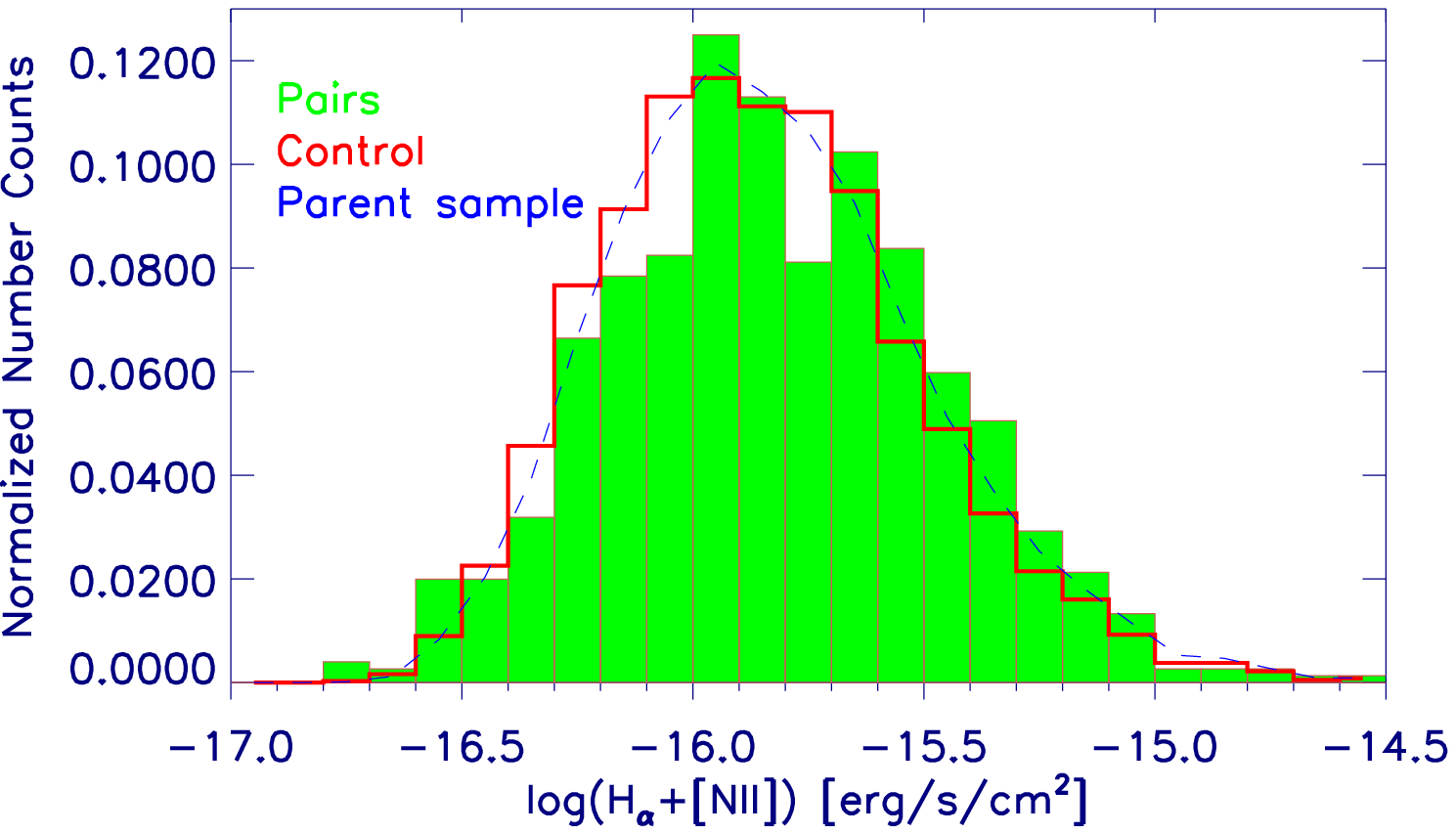} 
\end{center}
\figcaption{Redshift (top), absolute H band magnitude (K-corrected, middle), 
and raw \halpha\ flux (bottom)
distributions of the ELG pairs (solid green), 
the control sample of isolated ELGs (red),
and the parent ELG sample (blue dashed line).
The increasing difficulties of line identification explain the steep drop at z $> $1.5, 
where the \halpha\ line falls out of the wavelength coverage;
while the lower S/N at the G102/G141 overlapped region and misidentification of single-line emitters
contribute to the drop at $z=$0.8-1.1.
The K-S test of the similarity between the pair and control sample for the redshift, M$_{\rm H}$, 
and \halpha\ gives probabilities of 0.79, 0.005, and $<$1e-4, respectively.
\label{fig:zdist}}
\end{figure*}

A control sample of 4070 ELGs was then selected 
from the isolated ELGs to have the same redshift distribution as the pair sample, 
also requiring at least 
an \halpha\ (\nii\ corrected, see \S\ref{sec:n2}) 
or \oiii\ emission line flux detected with S/N $>$3.
The selection was done by choosing 10 isolated galaxies for 
each ELG pair system in the same redshift bin of the primary member, with a bin size of 0.1,
with no duplications.
If fewer than 10 galaxies are available,
we just use those. 
The redshift distributions of the control and pair samples are required to be identical by design, as confirmed by 
Kolmogorov-Smirnov (K-S) test probabilities of $p=$ 0.79, 0.96, and 0.96 that there are
no differences,
for the combined $z$ distribution, 
and the individual $z$ distribution for the primary and secondary pair members, respectively.
Figure~\ref{fig:zdist} (top) shows the redshift distribution 
ELG pair sample (413) and the control sample of isolated ELGs (4,070). 
and the parent ELG sample (8,192).

The pair and control samples show a generally comparable redshift distribution,
with two broad peaks around $z=0.5$ and $z = 1.1$,
corresponding to the redshifts at which \halpha\ falls in the most sensitive wavelength 
range of the grism coverages (throughput $>$10\%, G102: 0.81--1.15 \micron, G141: 1.08--1.69 \micron). 
Compared to the parent ELG sample,
pairs are less often found at $z > $1.5 and 0.7 $ < z < $0.9,
due to the S/N requirement for both pair members.
The reasons are two-fold. 
First, at the intermediate redshifts of 0.7 $< z <$ 0.9,
where the two grisms overlap,
the spectra are typically noisier due to the reduced
sensitivities in the overlap regions.
This increases the chances of missed emission lines and misidentifications.
Given the requirement of line detection in both ELGs 
to make place them in the pair sample,
the ELG incompleteness is doubled for pairs, 
which contributes to the deeper drop in number distribution at 0.7$< z < $0.9. 

In addition, at $z >$ 1.5, as \halpha\ shifts beyond the red limit of the G141 grism coverage, 
the redshifts are identified by \oiii\ emission lines. 
Based on observations up to $z=1.5$, for the same object,
\oiii\ lines are generally weaker than the \halpha\ emissions,
thus more difficult to detect \citep{mehta15}. 
In fact, only 7\% of our pair sample are [OIII]-only pairs. 
According to simulation results,
in the case of a single emission line,
about 6\% of the \halpha\ lines could be misidentified as \oiii\ lines \citep{colbert13}.
On the other hand, single-line \oiii\ emitters are rare, 
as they are often show a marginally resolved blue wing from the weaker doublet line, and are often accompanied by \hbeta\, making them unlikely to be misidentified \citep{baronchelli20}.
In the $z$ range of 0.7--0.9, about 29\%   
of the systems are single-line systems, 
higher than the average value of 15\% at all redshifts.
The combination of noisier spectra and single emission line identification
contributes to the ELG number drop in this redshift range.

In addition, we also compare the k-corrected absolute H band magnitude
and the \halpha$_{\rm raw}$ flux distributions in Figure ~\ref{fig:zdist} (middle and bottom). 
Since we do not have the mass measurements for the whole galaxy sample, 
we instead use the absolute H band magnitude as a proxy for the stellar mass. 
Despite their comparable median and standard deviation:
 -21.9$\pm$1.5 for pairs; -21.6$\pm$1.7 for the control; 
the K-S test probability is 0.005 for the the absolute H band magnitude comparison, 
indicating intrinsically different distributions. 
The pair sample shows a higher fraction of luminous members at $M_{\rm H} <$ -22. 
This is a selection effect related to both the lack of enough 
luminous isolated ELGs at $M_{\rm H} <$ -22
and the bias against fainter ELG pairs,
where both members are required to have at least one ELG detected at $>\,3\sigma$.
The \halpha$_{\rm raw}$ distribution also differs between the pair and control samples,
with the ELG pair sample showing higher median flux: 
(1.45$\pm$3.31)\,$\times10^{-26}$ \ergs\,cm$^{-2}$ for pairs; 
(1.31$\pm$3.57)\,$\times10^{-26}$ \ergs\,$cm^{-2}$ for the control,
where the comparable errors show the standard deviations.  
This is consistent with the SFR enhancement reported later 
in Sec~\ref{sec:sfr}.

\subsection{[NII] Correction}
\label{sec:n2}
Since the WFC3 grism spectra do not resolve 
the \nii\ $\lambda$6548$+$6583 doublet from the \halpha\ emission,
we needed to apply a correction to the \nii-blended \halpha$_{\rm raw}$.
In earlier WISP studies,  
either a uniform average flux correction of 29\% was applied \citep{colbert13},
or mass-dependent binned correction values ranging from 4.5\% to 19.5\% were used \citep{dominguez13}. 
Based on high-resolution Magellan/FIRE spectra of individual WISP galaxies at $z < $1.5, 
flux corrections from 6.4\% to 39.7\% were measured in \citet{masters14}
with an average of $\sim$17.5\%.  
As the \nii\ correction is found to be redshift- and mass-dependent \citep{erb06, sobral12, masters14}, 
the \halpha\ fluxes corrected with a uniform value may be underestimated
for the less-massive sources
and overestimated for the more massive galaxies.

\begin{figure*}
\begin{center}
\includegraphics[scale=0.7]{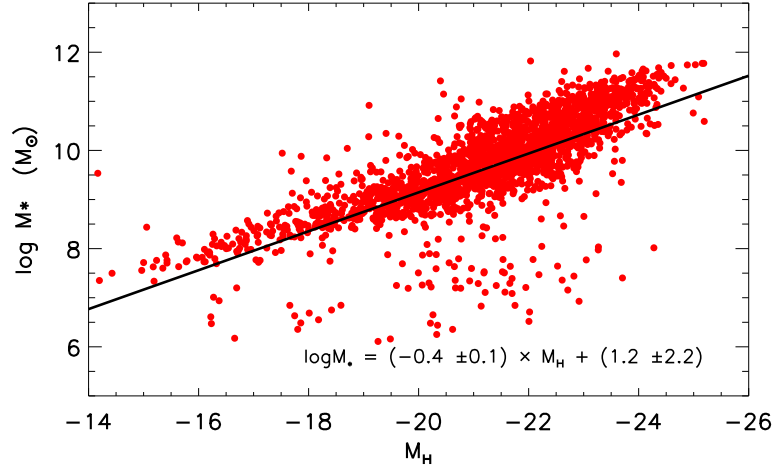} 
\end{center}
\figcaption{The distribution of the subsample of 2205 ELGs 
with stellar mass estimates from CIGALE (colored dots)
as a function of the K-corrected absolute H band magnitude. 
The straight line shows the best-bit linear correlation between the stellar mass and $M_{\rm H}$, 
with the slope and interception listed in the legend. This corresponds to a directly linear proportionality between the infrared light and the stellar mass.
\label{fig:msmh}}
\end{figure*}

In this paper, 
we adopted the average of two different methods to correct the \nii\ from \halpha$_{\rm raw}$,
similar to \citet{dominguez13} and \citet{atek14}. 
We first applied the redshift- and stellar mass- dependent 
correction function from \citet{faisst18}.
For galaxies with no stellar mass estimates,  
we used an empirical conversion of log$M_{*} = \kappa \times M_{\rm H} + c$,
where $M_{\rm H}$ is the K-corrected absolute H band magnitude,
$\kappa = (-0.4\pm0.1)$, and $c = (1.2\pm2.2)$ (Figure~\ref{fig:msmh}). 
These values are derived from the 27\% WISP ELGs (2205/8192) with     
$M_{\rm H}$, 
IRAC coverage 
and SED-based mass measurements.
 The [NII] corrections are 9\%,16\%, and 33\%
in the 3 mass bins:
log$(M_{*}/\msun) < $9,
9 $<$ log$(M_{*}/\msun) < $10,
and log$(M_{*}/\msun) > $10. 
Given the high model-dependency of the mass estimates,
we also derived the \nii/\halpha\ ratio from equivalent width of EW(\halpha$_{\rm raw}$). 
This is based on the the empirical relation from \citet{sobral12}. 
The corresponding [NII] contributions to the total line blend are 11\%,14\%, and 21\% in the 3 mass bins, respectively.
Thus averaged correction factors of 10\%,15\%, and 27\% were removed 
from the \halpha$_{\rm raw}$ flux in the corresponding mass bins. 

\section{Star Formation in ELG Pairs}
\label{sec:sfr}
In this section,
we will focus on galaxies with SFR estimates from the \halpha\ emission line measurements,
which are the majority of the ELG sample. 
In 93.5\% of the ELG pairs,
either one (2.0\%) or both (91.5\%) member galaxies have an \halpha\ detection.
In comparison, given their ELG nature,
93.7\% of the control group also have \halpha\ detections.

 We first perform the extinction correction for the \nii\-removed \halpha\ flux
based on the E(B-V) calculated from the observed Balmer decrements.
Since not all pairs in our sample have access to both \halpha\ and \hbeta\ lines, 
we adopt the mean extinction values from all ELGs with \halpha/\hbeta\ values
in three mass bins of log$(M_{*}/\msun)  <$ 9, 
9 $<$log$(M_{*}/\msun) <$10, 
log$(M_{*}/\msun) >$ 10. 
The average E(B-V) in these bins are: 
[0.07, 0.06, 0.17]\,mag, respectively.   
These values are calculated following the reddening curve 
of A$_{H_{\alpha}}$ from \citet{calzetti00},
assuming an intrinsic \halpha/\hbeta\ ratio of 2.86 for Case B recombination \citep{osterbrock89}. 
Our results are consistent with the 
values derived in \citet{dominguez13} for similar WISP ELGs at 0.75 $< z < $1.5,
\citet{atek14} for ELGs at 0.3 $< z < $2.3
and in \citet{momcheva13} for ELGs at $z\sim$0.8.
These extinction values are then applied to each stellar mass bin
to correct for the dust attenuation.  

The SFR was calculated based on the \citet{kennicutt98} relation (corrected for extinction),
assuming a Salpeter IMF:
\begin{equation}
SFR (\msun\,{\rm yr^{-1}}) = 7.9 \times\,10^{-42}\,L_{H_\alpha} (\ergs)
\end{equation}
where $L_{H_\alpha}$ is the luminosity of the \nii\ and dust extinction corrected
 \halpha$_{\rm raw}$ emission line.
We then divided the SFR by a factor of 1.8 to match the Chabrier IMF\citep[e.g.][]{gallazzi08}.  
For both pair and control samples,
we limit our SFR analysis to systems with \halpha\ measurements for a more reliable 
SFR estimate.
This rejects 6.5\% of the pair sample and 7\% of the control sample, 
where only \oiii\ is available.
 Since we rely on \halpha\ for the SFR estimates,
and base the following discussion only on the comparisons between 
pairs and the control sample of isolated ELGs,
the AGN contribution is ignored the following discussions. 
As we will discuss later in $\S\ref{sec:frac}$,
comparable AGN fractions are found in the pair and control samples. 

\begin{figure*}
\begin{center}
\includegraphics[scale=0.7]{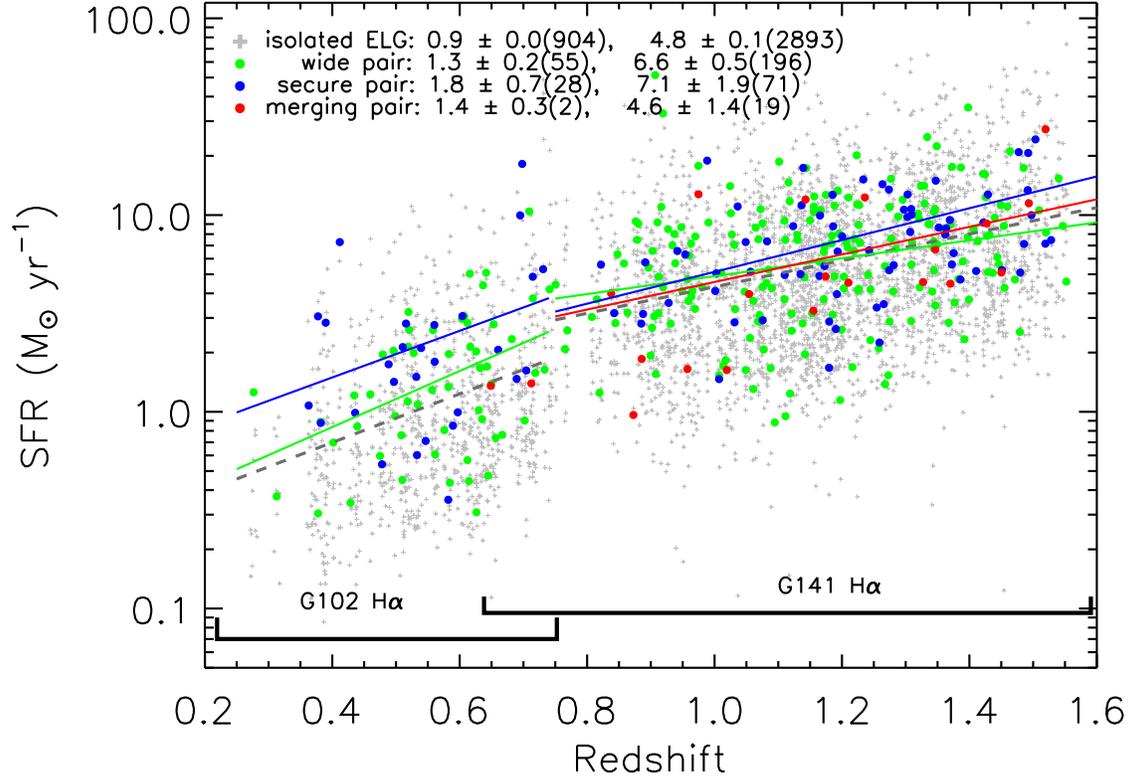} 
\end{center}
\figcaption{The redshift-SFR relation for in ELG pairs (colored dots)
and the control sample of isolated ELGs (grey crosses).
SFR are derived from \halpha\ 
emission line fluxes, corrected for \nii\ contamination and dust extinction.
The green, red, and blue dots mark the SFR in the wide, secure, and merging pairs in our sample.
The corresponding lines are the z $vs$ log(SFR) linear fit for each category, weighted by SFR uncertainties
and binned in two redshift ranges ($z <$ 0.75 \& $z > 0.75$), corresponding to G102 and G141 coverages. 
In the legend, we show the median and standard deviation of the SFR in each bin,
with the number of objects in the brackets for each bin. 
Due to small number statistics, no fit was given for the merging pairs at low-$z$. 
Except for the merging pairs,
average SFR enhancements of 1.4\,$\times$ - 2.1\,$\times$
are observed between the ELG pair sub-samples 
and the control sample of isolated galaxies.
\label{fig:sfrz}}
\end{figure*}

We compare the distributions of extinction-corrected SFR estimates
for the ELG pairs to the control sample. 
Figure~\ref{fig:sfrz} shows the SFR vs redshift distribution,
divided into two redshift bins below and above $z=$0.75, where the 2 grisms overlap.
For each subsample,
the errors are the standard deviation from the IDL curve-fit assuming the power-law function:
\begin{equation}
log (SFR) =  p0 + p1\times\,(z)
\end{equation}
where p0 and p1 are the intercept and slope of the power-law fit,
and the input SFR is weighted by the S/N of the \halpha\ emission line.
Overall, the pair sample shows marginally elevated SFRs 
with respect to the isolated ELGs. 
Compared to the control group,
the median SFRs for the subsample of wide, secure, and merging pairs
are enhanced by (1.5$\pm$0.3)$\times$, (2.1$\pm$0.8)$\times$, and (1.6$\pm$0.4)$\times$ at $z < 0.75$
and by (1.4$\pm$0.1)$\times$, (1.5$\pm$0.4)$\times$, (0.9$\pm$0.3)$\times$ at $z > 0.75$, respectively. 
Across all redshifts, we find an average enhancement of (40$\pm$20)\% in pairs. 
Table~\ref{tab:sfr} summarizes the enhancement values for different subsamples of the ELG pairs. 

We then consider the major pairs in our sample (i.e. $\sim$70\% of our sample, 
based on H band flux ratios).
Compared to isolated ELGs,
the median SFR for major pairs show
an average enhancement of (2.1$\pm$0.5)$\times$, and (1.4$\pm$0.1)$\times$ at low and high z, respectively.  
These enhancement could be further broken down into:
(2.3$\pm$0.5)$\times$, (2.4$\pm$1.1)$\times$, (1.6$\pm$0.4)$\times$
and (1.5$\pm$0.1) $\times$, (1.5$\pm$0.2) $\times$, (0.9$\pm$0.3)$\times$
for the $z< 0.75$ and $z > 0.75$ bins.
Compared to all pairs, the major pairs show a marginal increase in the level of the SFR enhancement,
especially at $z<$0.75 for major pairs in the wide and secure subsamples,
but not in the merging samples, which has limited statistics.
Across all redshifts, the average enhancement in major pairs are (60$\pm$20)\%. 

Next, we only consider pairs with \voff $ < \,300\kms$, 
which are the systems most likely to be undergoing interactions. 
About 25\%, 60\%, and 62\% of the   
wide, secure, and merging pairs meet this stricter \voff\ selection.
Pairs closer in physical space are more likely to be associated. 
The average enhancements at low and high $z$ are
(2.1$\pm$0.5)$\times$ and (1.5$\pm$0.1)$\times$, respectively.
Their SFR enhancements are
(2.5$\pm$0.4)$\times$, (2.4$\pm$1.5)$\times$, and (1.6$\pm$0.3)$\times$ at $z<$ 0.75,
and (1.4$\pm$0.2)$\times$, (1.5$\pm$0.6)$\times$, and (0.9$\pm$0.2)$\times$ at $z>$ 0.75, respectively.
After applying the \voff\, constraint, 
the enhancement at lower $z$ becomes more significant,
while at higher $z$ the SFR enhancement remains almost the same 
with and without any additional selection criteria.
This results in an average SFR enhancement of (65$\pm$20)\%.

Overall, these results are comparable to what was found in local SDSS pairs, 
where SFR enhancements of 1.2$\times$--1.5$\times$ were observed 
with a 10-20\% uncertainty 
at separations between 150 --30\,kpc \citep[e.g.][]{patton13}, 
corresponding to the wide and secure pairs in our sample.
Our slightly higher enhancement in the secure pairs 
may be related to the ELG nature of this sample,
when both members have to satisfy the ELG selection. 
The enhancement is also more significant among the lower-$z$ pairs, 
especially when only considering the major or \voff $ < \,300\kms$ subsamples.  
This is consistent with the results of \citet{xu12b}.
The much higher enhancement (2$\times$--3$\times$) found in SDSS for their \dproj$<$10\,kpc pairs
is not seen here in our merging pairs (0.9-1.6$\times$ enhancement), 
possibly related to our small number statistics.  
 Although dust extinction is corrected in our analysis,
we note that if merging pairs suffer higher dust extinction than isolated galaxies,
the average correction applied to the SFR, based on all ELGs in the mass bins,
may still be significantly underestimated in merging pairs.
It is worth noting that we are comparing the SFR 
in ELG pairs and isolated ELGs.
If quiescent galaxies were also included in the control sample, 
the actual enhancement is likely to be higher than what is reported here. 

Next, we make the SFR comparison between normal pairs and the pairs with disturbed morphology.
We find enhanced SFR in the 112 pairs showing tidal tails or disturbed 
morphology--likely in the process of merging.  
At the low and high $z$,
the disturbed pairs show SFR enhancement of (1.9$\pm$0.5) and (1.5$\pm$0.1) $\times$ over the isolated ELGs, 
but no more than a marginal SFR increase compared with the compact ELGs (1.1$\pm$0.3 and 1.1$\pm$0.1, respectively).

\begin{figure*}
\begin{center}
\includegraphics[scale=0.8]{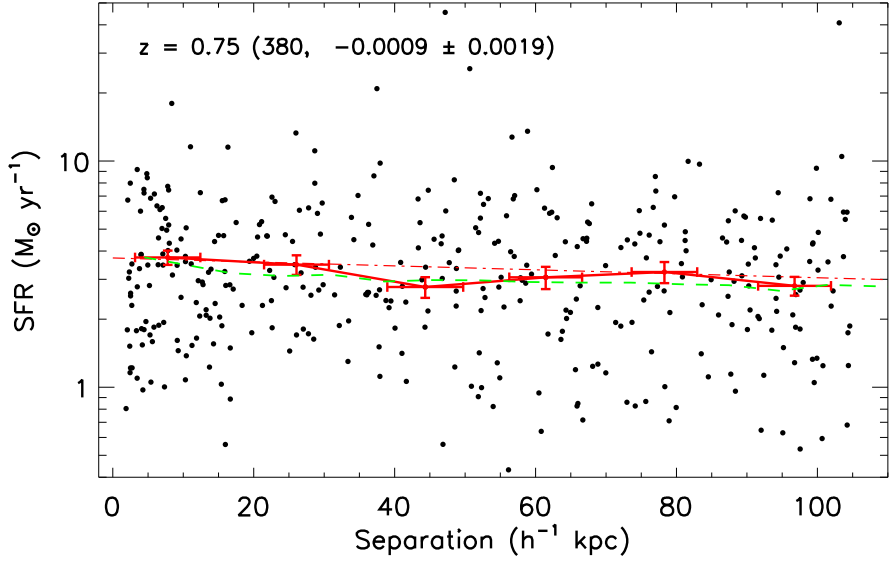} 
\end{center}
\figcaption{
The average SFR in ELG pairs as a function of pair separation,
normalized to a redshift of $z=$0.75. 
The number of objects and the slope are marked in the legend. 
The dash-dotted line marks the linear fit between the separation and log$(SFR)$.
The observed negative slope is consistent with 
slightly enhanced SFR towards smaller separation.
The median SFRs and corresponding mean standard deviations 
in bins of separations are marked in red crosses.
The dashed green line marks the relation for local SDSS pairs at $z<$0.2 \citep{patton13},
normalized to our data point at $\sim$100\,h$^{-1}$kpc.
Despite the large uncertainties, 
the binned SFR shows a general trend of increasing towards smaller separations, 
especially between $\sim$50 and 5\,h$^{-1}$kpc. 
The level of SFR enhancement in ELG pairs is consistent with the local relation for SDSS pairs. 
\label{fig:sfrd}}
\end{figure*}

We then study the SFR enhancement as a function of pair separations (Figure~\ref{fig:sfrd}).
To minimize the redshift effect, we normalize the SFR to $z=0.75$, 
by adopting the average linear correlations for the pair sample in Figure~\ref{fig:sfrz}:
\begin{equation}
log (SFR) =  (-0.37\,\pm\,0.78) + (1.16\,\pm\,1.32)\times\,z     
\end{equation}
for $z <$ 0.75, and
\begin{equation}
log (SFR) =  (-0.04\,\pm\,0.14) + (0.64\,\pm\,0.40)\times\,z    
\end{equation}
 for  $z \ge$ 0.75.
In Figure~\ref{fig:sfrz}, 
an insignificantly negative slope (-0.0009\,$\pm$\,0.0019) is observed for the full ELG pairs sample.  
After binning the data by separations (Figure~\ref{fig:sfrd}, red crosses),
the marginal increase of SFR towards smaller separation
is confirmed, especially from $\sim$50\,h$^{-1}$\,kpc:
the SFR is on average 25-35\% higher at 20--5\,h$^{-1}$\,kpc than at $\sim$45\,h$^{-1}$\,kpc,
though at a $<\,1\sigma$ level.
Increased SFR at low pair separation is often associated with 
interaction-triggered star formation, as observed in local galaxies (Figure~\ref{fig:sfrd}, green dashed line).
The SDSS relation is also plotted at $z = $0.75 (green curve),
normalized to the binned SFR at $\sim$100\,h$^{-1}$\,kpc,
for a better comparison with our sample.
The amount of SFR increase is comparable: 
25-43\% from $\sim$45\,h$^{-1}$\,kpc to 5--20\,h$^{-1}$\,kpc.
Further out, SDSS sees a SFR enhancement of 
$\sim$10\% from $\sim$100\,h$^{-1}$\,kpc to 45\,h$^{-1}$\,kpc,
and the ELG pairs also show a $\sim$10\% increase in the same distance range. 
In brief, 
the ELGs pairs show a weak increase in SFR towards smaller separation,
similar to what was found locally with the SDSS pairs. 

\section{Emission Line Ratios}
\label{sec:stack}
Our grism spectra enable the study of 
emission line properties over a range of galaxy parameters 
including separations (pair type), redshift, line equivalent width, and star formation rates.
In this section we use spectral stacking to 
explore the variations of the emission line ratios in various subsamples of ELG pairs. 
Only galaxies with both G102 and G141 coverage are included
in the following analysis.
This applies to 75\% of the pair sample (309 pairs) and 72\% of the control sample (2934 galaxies).  

We adopt a stacking procedure as detailed in \citet{henry13} and \citet{dominguez13}.
In brief,
after masking out the emission-line regions,
we first subtract a model continuum for each galaxy in the rest-frame.
All spectra are visually inspected to make sure the subtraction is properly carried out.
This is done by smoothing with a 20-pixel boxcar for the
G102 (dispersion of 24.5 \AA\,pixel$^{-1}$)
and a 10-pixel boxcar for G141 (46.5 \AA\,pixel$^{-1}$).
To have an equal contribution from all galaxies,
we normalize each spectrum by selected emission line flux,
measured from fitting a single Gaussian profile in the line region.
As will be described below, 
\halpha\ and \oiii\ fluxes are used for normalization, respectively,
in the selected redshift bins for stacking. 
In the case of high $z$ sources (e.g. \oiii/\oii (3727\AA) ratios, see below),
where \halpha\ falls out of the spectral coverage,
\oiii\ flux is used for the normalization.
Then, after combining the normalized spectra,
we use the median value to generate the stacked spectrum. 
Finally, in each bin,
we use the bootstrap method to resample the input spectra,
estimating the errors from the RMS (root-mean-square) of 500 artificial stacks. 

\begin{figure*}
\begin{center}
\includegraphics[scale=0.9]{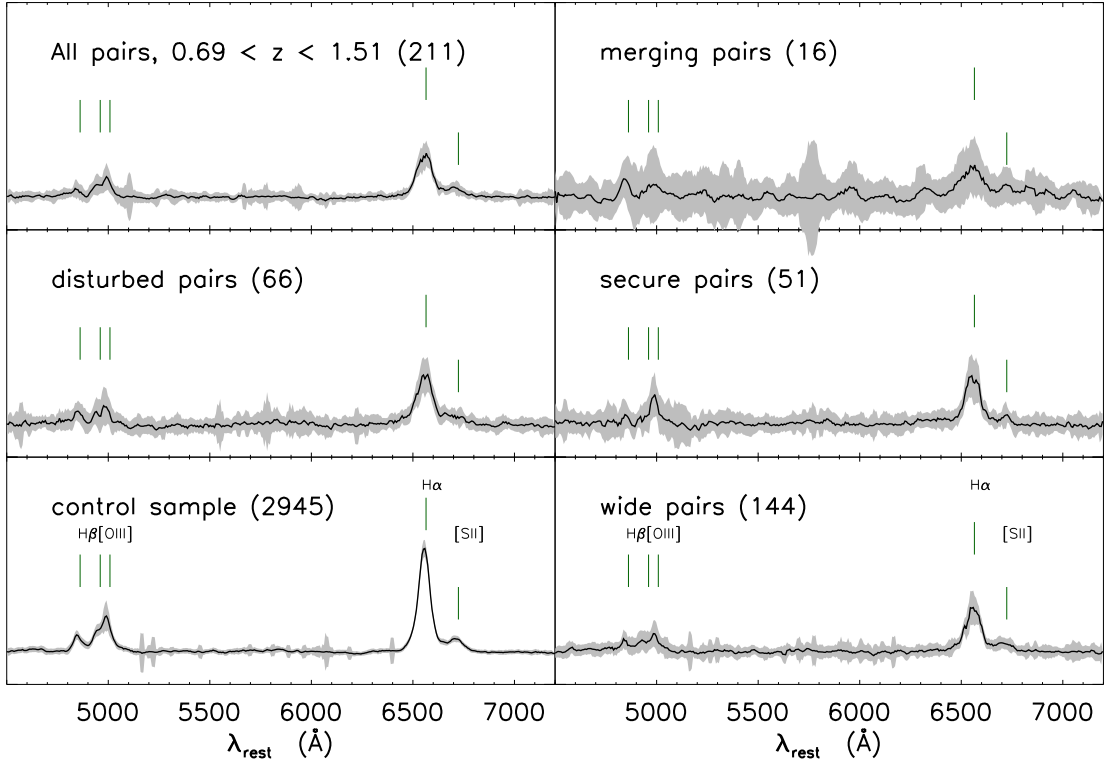} 
\end{center}
\figcaption{Stacked median spectra (continuum subtracted) for ELG pairs in bins of pair types, 
with 1$\sigma$ errors in grey. 
In brackets are the numbers of spectra included in each stack. 
Only pairs with certain $z$ range and full grism coverages are included, 
to ensure simultaneous \halpha\ and \oiii\ coverage.
The three right panels are the subsets of `all pairs' (top left panel), 
while middle left panel (`disturbed') include all the disturbed pairs regardless of 
their separation and velocity difference. 
\label{fig:stack}}
\end{figure*}

Figure~\ref{fig:stack} shows examples of the stacked median spectra for our pair subsamples
in selected bins.
For comparison we also plot the stack of the control sample of isolated ELGs at the same redshift range (bottom left).  
We only compare the stacked spectra in the range of 0.69 $ < z < $1.51, where both \halpha\ and \oiii\
lines are covered. 
The emission line ratios of \halpha\ to \oiii\ are then calculated
by fitting the spectra, allowing up to three Gaussians per line,
to allow multiple velocity components, to account for 
the possible contribution of \nii\ and the weaker \oiii\ doublet, 
which is fixed at one third of the flux of the stronger \oiii\ doublet.  
We repeat the same procedure  
to calculate other line ratios (\oiii/\oii, \sii/\oii, \halpha/\hbeta) in 
different redshift ranges,
and in different SFR and EW bins.
Here the mean SFR 
of the paired galaxies is used in the three SFR bins;
and for the EW bins, we require both pair members to satisfy the requirement to be included in the stack. 

Table~\ref{tab:ratio} summarizes the relative line ratios from the stacked spectra
in different redshift bins, with S/N$>$3 ratios marked in boldface.  
The pairs are divided into three redshift ranges,
where these lines are covered:
0.69 $< z_1 < $ 1.51 (\hbeta, \oiii, \halpha, \sii), 
1.28 $< z_2 < $ 1.45 (\oii, \halpha, \sii), 
and 1.28 $ < z_3 < $ 2.29 (\oii, \oiii). 
No ratio is recorded if the stacked spectra is missing certain lines or 
is too noisy (S/N $<$ 1). 
We note that the stacked flux ratios may be inconsistent
with the individual measurements, due to a combined effect of 
bias by the non-detection of the emission lines---other than the \halpha\ or \oiii\ lines used for normalization;
universal [NII] correction for the \halpha\ flux---instead of mass-independent correction in the individual pairs;
and the lack of dust extinction correction for the stacked spectra---otherwise applied to individual spectra.
Therefore, the absolute values of the stacked line ratios should be used with caution. 
In the following discussion, we only
focus on the trends of the line ratios presented in the stacked spectra.

Various factors such as gas density, metallicity, ionization parameter, and ionizing spectral index 
could each influence the observed line ratios in different ways \citep{yan12}. 
First of all, compared to the control sample, the overall higher SFR (See \S\ref{sec:sfr})
found in pairs is reflected in their relative line ratios,
which indicate that
pairs may have lower ionization levels or higher metallicities (Table~\ref{tab:ratio}).
We notice that the \halpha/\oiii\ ratio tends to be higher in pairs
than in the control sample, as shown in Figure~\ref{fig:zdist}.
This would be consistent with an enhanced star formation and possibly higher metallicities, 
which could be related to the two times larger masses of the pair galaxies.
As the SFR increases from low SFR (1-10\,$\msun\,yr^{-1}$) to high SFR ($>10\,\msun\,yr^{-1}$), 
we find a significant increase in the \halpha/\oiii\ ratio. 
Disturbed and merging pairs also show
marginally higher \halpha/\oiii\ ratios, 
consistent with enhanced SFR in these systems. 
On the other hand, the low \halpha/\oiii\ ratio in the high-EW bin 
is due to their noisy stacked spectra, 
where the \oiii\ line is blended in with the \hbeta\ line, 
yielding a higher \oiii\ flux with very low S/N. 
For the medium SFR bin, the lower ratio is real
and caused by the relatively weaker \halpha$+$[NII] line.   
In comparison, the \halpha/\sii\ ratios are generally lower in pairs than in the control sample,
suggesting an overall stronger \sii.
One possibility is shock excitation powered by tidal interaction in pairs,
which could generate strong low-ionization (nuclear) emission-line regions, i.e. LI(N)ER-like emission features \citep{monreal10, yan12,rich14, belfiore16}, 
and cause the relatively stronger low ionization lines that cannot be cancelled out 
by the SFR increase, for which \halpha\ is the proxy. 

Another possibility could be the
different abundances (metallicities) or excitation mechanisms, 
such as diffuse interstellar gas or change of hardness of ionization, 
which can also change the \halpha/\sii\ ratios. 
To test this, we estimate the metallicities for the pair samples based on the stacked spectra
using R23 \citep{kk04}.
Unfortunately, given the large uncertainties, 
the stacked metallicities
are not sensitive enough to show any significant difference between the control sample and 
the various pair subsamples (Table~\ref{tab:met}). 
The only exception is in the high SFR (SFR$>10\,\msun\,yr^{-1}$) bin,
which has a higher metallicity than the control sample. 
This is consistent with their relatively lower \oiii/\hbeta\ and \oiii/\oii\ ratios in Table~\ref{tab:ratio},
possibly related to higher masses in these high SFR objects.

\begin{figure*}
\begin{center}
\includegraphics[scale=0.7]{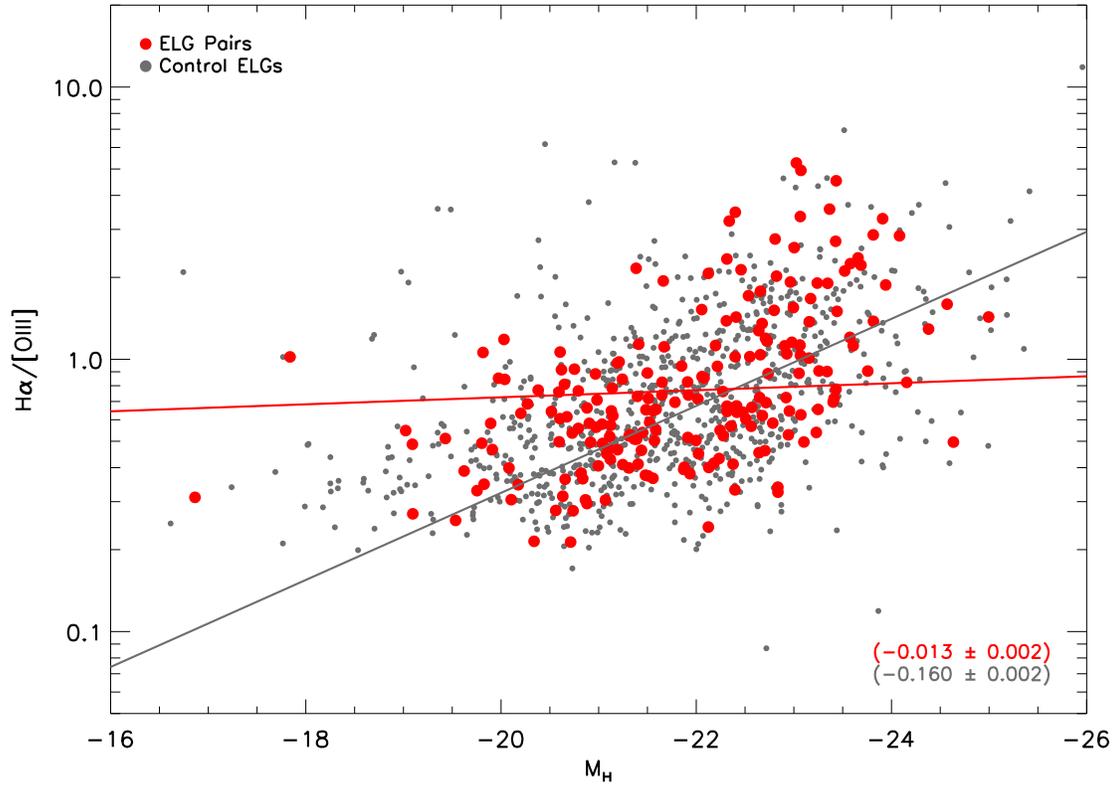} 
\end{center}
\figcaption{Distribution of the absolute H band magnitude ($M_{\rm H}$) versus the \halpha/\oiii\ line ratios
for the pairs (red) and control (grey) sample of ELGs.
Only pairs with S/N $>$ 3 for both lines are plotted. 
We find significant correlations in both pairs and the control sample (Spearman rank's probability P $\ll$ 0.001).
The correlation coefficients with respective errors between $M_{\rm H}$ and log(\halpha/\oiii) are given in the lower right corner. 
\label{fig:hao3}}
\end{figure*}

To test this, we compare the \halpha/\oiii\ line ratios with the absolute H band magnitude, $M_{\rm H}$, 
in Figure~\ref{fig:hao3}.
The \halpha/\oiii\ line ratios increase as the luminosity increases for both ELG pairs and the control sample.
The correlation is significant between $M_{\rm H}$ and \halpha/\oiii\ ratios, 
with Spearman rank probability P $\ll$ 0.001 for both samples, 
consistent with a decreasing excitation temperature of HII regions with increasing galaxy 
mass (with $M_{\rm H}$ as a proxy).  
This agrees with the strong anti-correlation 
between \oiii/\halpha\ ratio and B band luminosity found in \citet{moustakas06}.
The pairs sample shows a flatter slope,
suggesting lower excitation temperatures in 
low-mass galaxy pairs compared to single galaxies;
or higher excitation temperatures in more massive galaxy pairs.

The observed Balmer decrement, \halpha/\hbeta, 
a measure of dust attenuation,
has a large variation in different subsamples. 
We do not see any clear trend between the various
subsamples of ELG pairs, or between the ELG pair and the control sample. 
The only exception is that for high SFR pairs, 
the Balmer decrement is also higher, 
indicating more dust obscuration in galaxies with the highest level of star formation. 
Individual measurements, 
after dust correction and mass-dependent \nii\ removal,
show that a significant fraction ($\sim$40\%) of the pairs
that have both \halpha\ and \hbeta\ detected (SNR\,$>$\,1),  
having \halpha/\hbeta\ ratio greater than 2.86.
This is not surprising given the substantial amounts of reddening,
commonly seen in emission-line galaxies \citep{dominguez13, ly12}. 
The \oiii/\hbeta\ ratios in pairs 
agree within the errors with those of the control sample.  This is
consistent with the result from the BPT diagram described in Section~\ref{sec:frac}.

For $z>$1.28, when \oii\ emission lines are included,
pairs show lower \halpha/\oii\ and \sii/\oii\ ratios. 
Since the \oiii/\oii\ ratio is also generally lower in pairs, 
this indicates an overall strong \oiii\ but even stronger \oii\ in pairs. 
One possibility could be an overall lower metallicity
in pairs than in isolated ELGs, which however was not significant within our sample,
except for the high EW bins (Table~\ref{tab:met}).  

Compared to all pairs, the disturbed pairs have marginally higher \halpha/\oiii\ and \halpha/\sii\ ratios, 
as well as \halpha/\oii, \sii/\oii, and \oiii/\oii\ ratios, likely due to higher \oii\ thus stronger low ionization zones
in such systems. 
For the pair types from wide to secure to merging pairs, 
we observe no significant trend, given their large error bars.
The only exception is the increase of \halpha/\oiii\ ratio from secure to wide pairs,
which can be explained by a relatively weaker \oiii\ in wide pairs due to less interaction. 

In short,  we notice a general trend of weakly increased low ionization emission lines 
in pairs as compared to the control sample, 
although the degeneracy with the increased star formation, the uncertainties in line ratio and metallicity measurements,
and the resolution limits
make further quantitative analysis difficult at this stage. 
 
\section{ELG pair fraction and AGN fraction}
\label{sec:frac}
With this statistically significant sample up to $z\sim$1.5, 
in this section we try to calculate the ELG pair fraction, and study its evolution with redshift. 
The biggest challenge in calculating the ELG pair 
fraction is the complicated completeness correction associated with the ELG nature of our sample.
In this section, we will report the observed pair fractions,
after applying the completeness correction based mainly on the line flux, equivalent width,
and object size \citep{colbert13,bagley20}. 
Due to the flux limited nature of the WISP survey,
galaxies with weaker emission lines than the detection limit will be missed. 
As a result, an ELG with a faint companion 
is at greater risk of being missed 
than the brighter ELG pairs. 
We adopt the completeness corrections for the WISP ELGs (for details, see \citet{bagley20}, Appendix A1)
on an individual object-by-object basis, 
considering the emission-line equivalent width and `scaled flux' after adjusting for the depth differences,
as well as the object size and shape, which could also affect the line detectability of the grism spectra. 

Since the completeness was obtained for individual ELGs,
we first estimate the average completeness in the selected redshift bins. 
The binned completeness is calculated by dividing the total number of ELGs 
by total of the completeness (C) corrected numbers 
( i.e. $N_{\rm ELG} / \Sigma \frac{1}{C_i}$, 
where {C$_i$} is the completeness for the i$^{th}$ object out of the N objects in the bin). 
The true corrected ELG pair fraction is then estimated 
by applying the completeness of the primary member of the pair (C1),
the secondary member of the pair (C2),
and the total number of ELGs in the corresponding redshift bins (Ct):
\begin{equation}
f = \frac{N_{\rm pair}/(C1\times\,C2)}{N_{\rm ELG} / Ct}
\end{equation}
where f refers to the pair fraction\footnote{In groups of $\ge$ 3 members, 
only true pairs qualifying the selection in \S\ref{sec:sample} are counted, 
e.g. group of 3 members could be counted as 2 pairs or 3 pairs depending 
on their individual physical separations.}. 
Given the different flux values of the two members,
the completeness for the secondary pair member
is almost always lower than the primary pair.

We note that there are other factors,
besides the line identification and detection considered above,
which could also contribute to the completeness correction--
border effects (i.e. missing area close to the image boundaries);
decreased detector sensitivity at the blue ends of the spectra,
and the correction for selection effects due to mass ratios (major $vs$ minor), separations, 
velocity offsets or galaxy types (e.g. pairs made of ELG $+$ quiescent galaxy).
However, to convert observed pair fraction to a merger fraction, 
as often reported in the literature, 
requires assumptions and simulations of the evolution between ELGs and other galaxy populations. 
A fair comparison of the ELG fraction to merger rates reported for other galaxy populations (e.g. magnitude limited, color-selected, mass-selected samples), would 
require knowing the normalization of all the various selection effects
between the different samples,
which is beyond the scope of this paper. 
Therefore, in the following analysis, we will only focus on the ELG population,
rather than
extrapolating the ELG pair fraction to galaxy merger rate, 
or to comparing with other pair samples that were selected in different ways.

\begin{figure*}
\begin{center}
\includegraphics[scale=0.8]{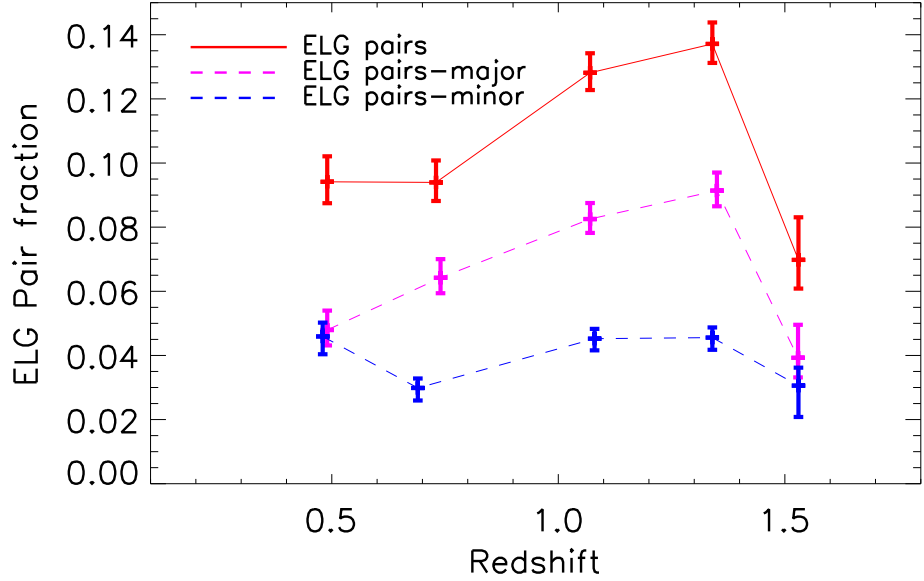} 
\end{center}
\figcaption{
Completeness-corrected ELG pair fraction as a function of redshift (with Poisson errors)
for the full pair sample (red), the subsamples of major (magenta) and minor (blue) ELG pairs,.
The drop at z $>$1.5 regime is 
highly uncertain due to the small-number statistics in that last bin.
At z $<$1.5, our result shows an increasing trend of the ELG pair fraction towards higher $z$,  
for both the full ELG pairs (red solid line), and the ELG-major pairs (magenta dashed line). 
The  ELG-minor pair fractions (blue dashed line), on the other hand,  
have a flatter trend along the redshift range covered. 
The data used to make this plot are listed in Table~\ref{tab:frac}.
\label{fig:frac1}}
\end{figure*}

The observed ELG pair fraction in our spectroscopically selected pair sample shows 
an overall increasing trend with redshift, 
after the completeness correction described above (Figure~\ref{fig:frac1}).
Except for the minor ELG pairs, 
we observe an increasing trend towards higher redshifts (Table~\ref{tab:frac}).

Unlike the full-sample and major pairs, 
the pair fraction for minor ELG pairs remains more or less flat over the observed redshfit range, although the drastic drop at $z>$1.5 is
highly  uncertain because of the small-number statistics with large uncertainties. 
Given the fainter nature of the secondary member of the minor pairs,
we are doomed to miss more minor pairs (i.e. in  which only the brighter one of the minor ELG pair is detected).
This incompleteness also gets more severe at higher redshifts for the same flux limit. 
Therefore the fractions shown here are only lower limits for the minor ELG pairs. 
We then fit the ELG pair fraction with a linear slope of $f \propto\,(1+z)^\alpha$ up to $z\sim$1.6,
a region with better number statistics, as shown in Figure~\ref{fig:zdist}. 
For the full ELG pair sample, $\alpha = {0.58\pm0.17}$,
while for the major and minor pair fraction (H band flux ratio defined, see \S\,\ref{sec:major}),
the linear fits show trends of $f \propto\,(1+z)^{0.77\pm0.22}$,
and $f \propto\,(1+z)^{(0.35\pm\,0.30)}$, respectively.
The full, major and minor ELG pair samples all have a positive $\alpha$,
indicating an increasing pair fraction with the redshift,
although for the minor ELG pairs,
the correlation is insignificant and consistent with being flat over the covered redshift range. 
We also note that, since some pairs may have one quiescent member (i.e. without emission lines), 
the ELG pair fraction reported here is only a lower limit for the ELGs that would eventually merge with another galaxy. 
While on the other hand, including quiescent galaxies would also increase 
the number of the parent sample; thus the direction of the fraction change cannot be easily predicted.
The intrinsic selection differences between the ELGs (i.e. no continuum required),
and other photometry- or mass-selected samples in the literature (i.e. photometry in broad bands required),
make it misleading to directly compare the various pair and merger fractions. 
For instance, in photometrically-selected pair samples,
the contamination of galaxies due to redshift uncertainties is less of an issue in the ELGs.

\begin{figure*}
\begin{center}
\includegraphics[scale=0.55]{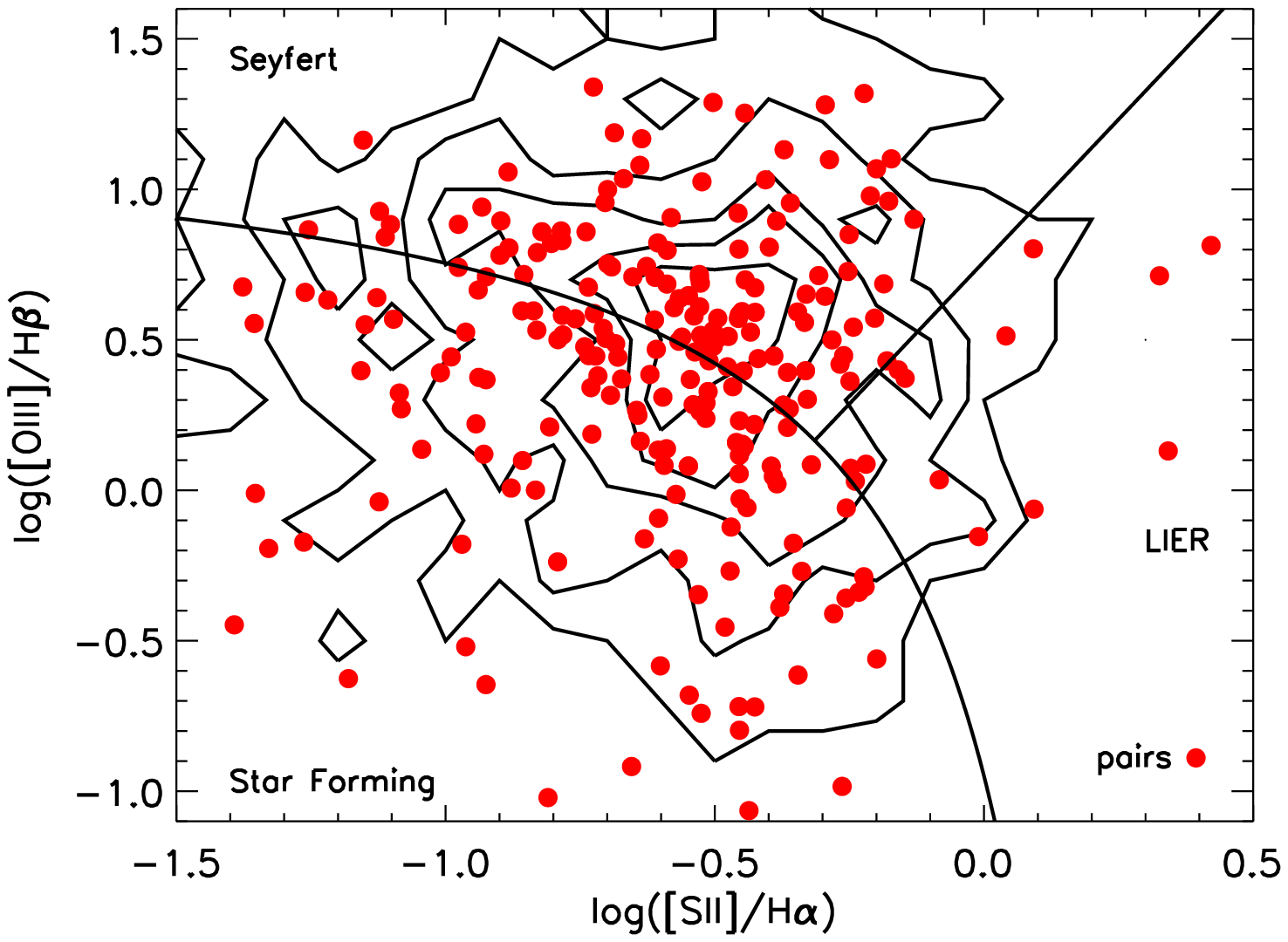} 
\includegraphics[scale=0.55]{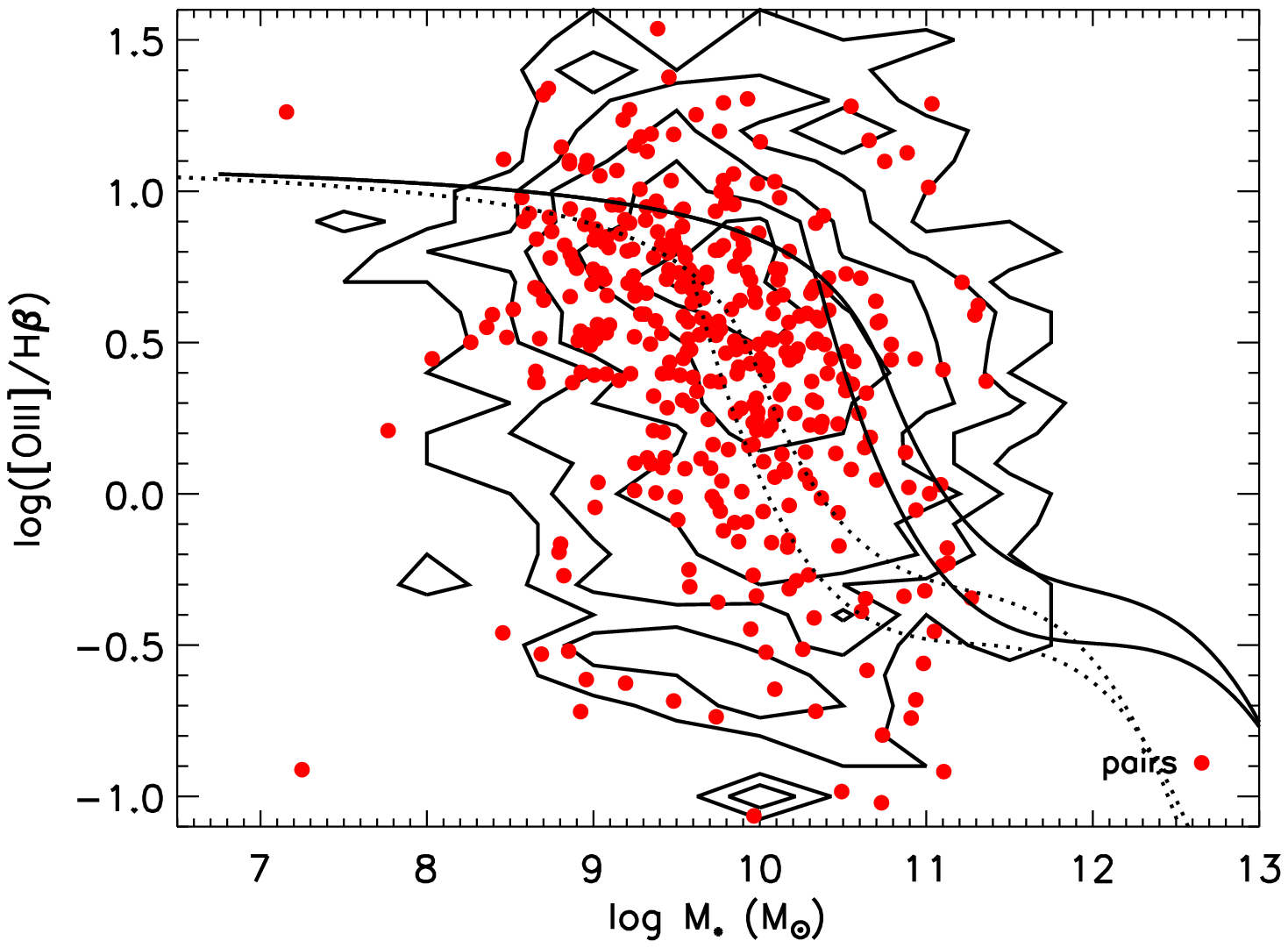} 
\end{center}
\figcaption{The distribution of the ELG pairs (red dots) 
and control samples (contours) in the modified-BPT diagram \citep[Left, ][]{bpt81,kewley06}
and in the Mass-Excitation (MEx) diagram \citep[Right,][]{juneau11, coil15}. 
The dividing curves are from \citet{kewley06} and \citet{juneau14}. 
No significant enhancement of AGN fraction is found in the pairs, 
with comparable fractions of $\sim$45\% in both samples. 
This fraction was also $\sim$40-45\% in the MEx diagram before the correction (dotted curves), 
and drops to $\sim$18-19\% after applying the offset (solid curves) from \citep{coil15} . 
\label{fig:bpt}}
\end{figure*}

Finally, we estimate the AGN fractions in pairs.
Based on numerical simulations, AGN activity can be triggered by the interaction of galaxies \citep[e.g.][]{hopkins06}.
Observational evidence of higher AGN fractions in pairs has also been found
in some studies, with a factor of 2-7 enhancement from $z\sim$0--1 \citep[e.g.][]{ellison11, goulding18}. 
Recent work using the COSMOS and CANDELS pairs, 
however, found no significant enhancement of AGN fraction in pairs \citep{shah20}. 
Given the ELG nature of our sample, 
we cannot directly compare the AGN fraction with other studies, 
which usually are not limited to certain galaxy types like ELGs. 
We can, however, make a comparison of the AGN fraction in ELG pairs and the control sample of isolated ELGs. 
Since the grism resolution does not permit \halpha-\nii\ separation, 
we chose the `modified-BPT' diagram   
using \sii/\halpha\ and \oiii/\hbeta\ \citep{bpt81, osterbrock85, kewley06}.
Here the blended \halpha$+$\nii\ is corrected for with mass-dependent corrections (See \S\ref{sec:sfr}). 
Figure~\ref{fig:bpt} shows the distribution of the subsample of ELGs 
with all four lines detected.
Unlike some of the earlier studies, we find no significant difference 
in the AGN fraction between pairs and the control sample galaxies. 
This may be related to the ELG nature
of our sample.
Considering all ELGs with BPT diagnostics without any modification,
the AGN fraction for the ELGs with high S/N lines (all 4 lines with S/N $>$3), 
though suffering from small number statistics and biases,  
is consistent within the errors: (33\%$\pm$19\%) for pairs 
and (43\%$\pm$10\%) for the control sample, respectively.  
 We note that these AGN frequencies are likely too high, because 
the dividing line between AGN and star-forming galaxies evolves between $z\sim$1 and $z\sim$0. 
At higher redshifts ($z > 1$),
the increased \oiii/\hbeta\ ratios
in star-forming galaxies would lead
to their being mistaken for Seyfert galaxies
if the local BPT relations were used
(see for example Henry et al. 2021
2021arXiv210700672H2021/07
The mass-metallicity relation at z~1-2 and its dependence on star formation rate
Henry, Alaina; Rafelski, Marc; Sunnquist, Ben and 24 more).
To compensate for this evolution, 
we test the AGN fractions based on the MEx selection \citep{juneau11, juneau14}. 
Here the stellar masses are estimated from $M_{\rm H}$ or CIGALE, 
if IRAC data are available (See \S\ref{sec:n2}). 
To reduce the contamination from high-$z$ star-forming galaxies,
we further apply the simple offset of 0.75 dex in stellar mass, following \citep{coil15}, 
to the \citet{juneau14} boundaries.
The AGN fraction is $\sim$18-19\% for both pairs and the control sample.
We note that before the correction,
the fraction based on MEx was also $\sim$40-45\%, 
similar to the local modified-BPT results.
As a third test, we match our sample to the ALLWISE\footnote{https://irsa.ipac.caltech.edu/} \citep{wright10} catalog,    
and find an AGN fraction (W1$-$ W2 $>$ 0.8) of 21$\pm$2\% for the control
and 22$\pm$5\% for the pairs.  Their AGN components are identified by the excess
thermal emission produced by hot dust in the W2 band.
Compared to AGN identification using ionized line ratios which may evolve from $z\sim\,0$ to $z\sim\,1$
and offset to higher values at higher redshift, 
using WISE colors may yield a more reliable AGN identification at higher redshift. 
Although all of the above methods have a high uncertainty, 
the lack of enhancement in pairs is confirmed. 
This is consistent with the results found in the COSMOS/CANDELS pairs,
with a larger sample  of more massive galaxies \citep{shah20}.

 In summary, 
despite the large uncertainties in AGN fraction identification, 
we observe no enhancement in the AGN fraction in pairs and the control sample of isolated ELGs.

\section{Summary}
\label{sec:summary}
By searching a total of 419 WISP fields with accurate emission line measurements,
we construct a statistically significant sample of 413 ELG pair systems, 
including 24 merging,
109 secure, 
and 281 wide pairs, 
according to our classification scheme ($\S$\ref{sec:sample}).
The ELG pair sample includes 47 higher-order systems with 3 or more members (11\% of the ELG pairs).
More than half ($\sim$63\%) of our pairs are at $z > 1$,
a redshfit range where spectroscopically-identified pairs were challenging to identify from ground-based observations.
The WISP survey contributes the largest spectroscopically-selected, 
unbiased galaxy pair sample at cosmic noon, 
countering the significant effects of cosmic variance that affects surveys limited to only small fields.

Compared to the control sample of isolated ELGs,
the ELG pairs show SFRs elevated by 40-65\% for various subsamples with different separations or velocity offsets. 
We observe a weak correlation between the SFR and the pair separation only at low redshift,
while at higher redshift ($z > $0.75), the correlation is flat, likely due to the large intrinsic scatter.
Despite the large uncertainties, after normalization to $z=$0.75, 
the ELG pair sample shows 
an increasing SFR at smaller pair separations, 
especially between $\sim$50 and 5\,$h^{-1}$\,kpc. 
The various line ratios based on our spectral stacking 
further indicate a general trend of slightly strengthened low-ionization lines in pairs. 
Finally, we study the ELG pair fraction ($f \propto\,(1+z)^\alpha$) 
and find an increasing power-law index of $\alpha\sim\,0.6$, 
though the uncertainties increase at higher redshift due to smaller number statistics,
yielding different $\alpha$ values for the full (0.58$\pm$0.17), major (0.77$\pm$0.22), and minor (0.35$\pm$0.30) ELG pair samples.
No enhancement in the AGN fraction is found in the ELG pairs 
as compared to the isolated ELGs. 
\\
\\
The authors would like to thank the referee for helpful suggestions. 
YSD thanks Andrea Faisst for helpful discussions. 
This research is based on observations made with the NASA/ESA Hubble Space Telescope obtained from the Space Telescope Science Institute, 
which is operated by the Association of Universities for Research in Astronomy, Inc., under NASA contract NAS 5-26555. 
These observations are associated with programs 11696, 12283, 12568, 12092, 13352, 13517, and 14178. 
Support for this work is also partly provided by the CASSACA
and Chinese National Nature Science foundation (NSFC) grant number 10878003. 
YSD acknowledges the science research grants from NSFC grants 11933003,
the National Key R\&D Program of China via grant number 2017YFA0402703,
and the China Manned Space Project with NO. CMS-CSST-2021-A05.
AJB acknowledges funding from the ``FirstGalaxies" Advanced Grant from 
the European Research Council (ERC) under the European Union¡¯s Horizon 2020 
research and innovation programme (Grant agreement No. 789056)¡±.
HA acknowledges support from CNES.

\begin{table}
\begin{center}
\caption{Emission Line Galaxy Pair Sample}
\begin{tabular}{lcccccc}
\hline
\hline
\vspace{3pt} WISPID         & R.A.(p)     & DEC(p)     & R.A.(s)     &  DEC(s)      & Sep    & $\Delta_V$ \\
 &&&&&  (\arcsec) & (\kms) \\
\hline
  1-10\_1-195 & 16.671364 & 15.151823 & 16.671064 &  15.1524   & 2.15 & 616.4    \\
  1-28\_1-124 & 16.634722 & 15.147898 & 16.634201 &  15.148602 & 2.72 & 667.8   \\
  5-41\_5-67  & 216.785309& 57.852722 & 216.784805&  57.850578 & 7.91 & 512.0  \\
\hline
\hline
\label{tab:sample1}
\end{tabular}
\end{center} 
Notes: R.A. and Dec are in J2000;  `p' in the bracket denotes the primary galaxy, `s' for the secondary galaxy. 
The typical redshift uncertainty is $\sim$0.1\% \citep{colbert13}, which translates to a velocity uncertainty of $\sim$70-200\,$\kms$,
depending on the actual redshift of the pair members. 
The full catalog is available in the online version of the paper. 
\end{table}

\begin{table}
\begin{center}
\caption{Galaxy Properties of the Emission Line Galaxy Pair Sample}
\begin{tabular}{lccccccccccc}
\hline
\hline
\vspace{3pt} WISPID        & z(p)      & z(s)     & Hflag & H(p) &  H(s)  & ${\rm H_\alpha}$ (p) & ${\rm H_\alpha}$(s) & SFR(p)  &  SFR(s)   \\
&& && (AB) & (AB) & $10^{-17}$ erg\,s$^{-1}$\,cm$^{-2}$ & $10^{-17}$ erg\,s$^{-1}$\,cm$^{-2}$ &  ($\msun\,yr^{-1}$)  & ($\msun\,yr^{-1}$)   \\ 
\hline
  1-10\_1-195 &  0.5084  & 0.5057   & 1 &  20.5 &   24.5   & 16.6$\pm$3.4 & 5.1 $\pm$1.5 &  2.2$\pm$0.4      & 0.5 $\pm$0.1   \\
  1-28\_1-124 &  1.3444 & 1.3396     &1 &  22.2 &    24.0  & 13.1$\pm$3.6 & 4.4$\pm$1.7 & 19.1$\pm$5.3     & 4.5$\pm$1.7      \\
  5-41\_5-67  &  1.3444  & 1.3481    &1 &  21.9 &    22.3   &17.6$\pm$1.3 & 7.2$\pm$1.8  & 25.6$\pm$1.9     &  10.6$\pm$2.6    \\
\hline
\hline
\label{tab:sample2}
\end{tabular}
\end{center} 
Notes: `p' in the bracket denotes the primary galaxy, `s' for the secondary galaxy.  Hflag: `0' for the default F160 filter, `1' for the F140 filter.  
The typical redshift uncertainty is $\sim$0.1\% \citep{colbert13}, which translates to a velocity uncertainty of $\sim$70-200\,$\kms$,
depending on the actual redshift. 
Here \halpha\ refers to the [NII]- corrected \halpha\ flux (see \S\ref{sec:n2}). 
SFRs are based on the [NII]-removed \halpha\ flux, and corrected for dust extinction (see \S\ref{sec:sfr}). 
The full catalog is available in the online version of the paper. 
\end{table}

\begin{table}
\begin{center}
\caption{SFR Enhancement in the Pair Samples}
\begin{tabular}{lcccc}
\hline
\hline
\vspace{3pt} Subsamples  &  \#   &   SFR Enhancement   &  \#    &    SFR Enhancement  \\
\hline
&  & $z < $ 0.75 & & $z>$0.75  \\
\hline
Wide    & 55 & 1.5$\pm$0.3 & 196 & 1.4$\pm$0.1  \\
Secure  &28 & 2.1$\pm$0.8 & 71 & 1.5$\pm$0.4 \\
Merging   &2  & (1.6$\pm$0.4)$^*$ &9 & 0.9$\pm$0.3 \\
\hline
&& Major Pairs && \\
\hline
Wide    & 30 & 2.3$\pm$0.5 & 133 & 1.5$\pm$0.1  \\
Secure  &20 & 2.4$\pm$1.1 & 49 & 1.5$\pm$0.2 \\
Merging   &2  & (1.6$\pm$0.4)$^*$ & 18 & 0.9$\pm$0.3 \\
\hline
&&$\Delta_V < 300 \kms$ &&\\
\hline
Wide    & 20 & 2.5$\pm$0.4 & 48 & 1.4$\pm$0.2  \\
Secure  &14 & 2.4$\pm$1.5 & 43 & 1.5$\pm$0.6 \\
Merging   &1  & (1.6$\pm$0.3)$^*$ & 12 & 0.9$\pm$0.2 \\
\hline
&& Morphology && \\
\hline
Disturbed  & 15 & 1.9$\pm$0.5 & 97 & 1.5$\pm$0.1 \\
Compact   & 64 & 1.7$\pm$0.4 &  164 &  1.3$\pm$0.2 \\
\hline
\hline
\label{tab:sfr}
\end{tabular}
\end{center} 
Notes: All enhancements are calculated with respect to the median SFR values for the control sample in the corresponding redshift bins. \\
$*$: Enhancement values calculated from the average SFR and their associated error. 
\end{table}

\begin{table*}
\begin{center}
\caption{Emission line flux ratios of the stacked spectra}
\begin{tabular}{ccccccccccc}
\hline
\hline
galaxy type  & N${_1}$ & \halpha*/\oiii\ & \halpha/\hbeta\ &  \halpha/\sii\  & \oiii/\hbeta  & N${_2}$  & \halpha/\oii\   & \sii/\oii  & N${_3}$ & \oiii/\oii \\
\hline
\hline
control sample      &  2945 &  {\bf 2.18 $\pm$  0.16} &   {\bf 6.58 $\pm$  0.84} &   {\bf 5.21 $\pm$  0.42} &   {\bf 3.02 $\pm$  0.49} &  455  & 4.31 $\pm$  2.72 &   1.30 $\pm$  0.97 & 758   & {\bf 3.16 $\pm$  0.31} \\
\hline
         all pairs          &  211  &  {\bf 2.55 $\pm$  0.52} &   8.09 $\pm$  2.98 &   {\bf 4.24 $\pm$  0.94} &   3.18 $\pm$  1.24 & 50 &   3.36 $\pm$  0.58 &   0.46 $\pm$  0.12  & 87  &     2.27 $\pm$  0.96  \\
             disturbed  & 66    & 3.55 $\pm$  1.20 &   6.41 $\pm$  2.52 &   {\bf 4.68 $\pm$  1.37} &   1.81 $\pm$  0.98  & 14 &  3.93 $\pm$  2.78 &   0.91 $\pm$  0.84 & 26  &  2.77 $\pm$  1.52 \\

  merging & 16    &  3.54 $\pm$  2.87 &   ...                          &   3.33 $\pm$  2.02 &   ...                         &  3   & {\bf 3.18 $\pm$  0.90} &   0.57 $\pm$  0.31 &  8&  {\bf 2.44 $\pm$  0.76} \\
   secure  &  51    & {\bf 1.98 $\pm$  0.59} &   9.89 $\pm$  8.47 &   6.23 $\pm$  3.15 &   4.99 $\pm$  4.63  & 10 & ...                        &     ...              & 24  &   3.44 $\pm$  1.97   \\
    wide   &  144 &  {\bf 2.91 $\pm$  0.71} &   6.08 $\pm$  2.09 &   {\bf 4.73 $\pm$  1.29} &   {\bf 2.09 $\pm$  0.90}  & 37 & 9.09 $\pm$  8.15 &   ...               & 55  &   1.81 $\pm$  0.94 \\
\hline
EW [\AA] &&&& \\
   $> 500$     &  23   &1.86 $\pm$  0.89 &   3.73 $\pm$  2.17 &   5.07 $\pm$  2.06 &   2.01 $\pm$  1.59 & 4 &  {\bf 1.51 $\pm$  0.04} &   0.03 $\pm$  0.02            & 11 & {\bf 3.24 $\pm$  0.82}    \\
   100 -- 500  & 146  &{\bf 3.41 $\pm$  1.15} &   7.55 $\pm$  4.08 &   {\bf 4.31 $\pm$  1.17} &   2.22 $\pm$  1.53 & 33 &     3.62 $\pm$  2.31 &   0.75 $\pm$  0.57 & 39 &  1.77 $\pm$  1.15        \\
   $< $100     &  8     &     ...   &      ...    &      ...    &     ...    & 2   &    ...  &       ...  & 17 &  1.52 $\pm$  0.56  \\
\hline
SFR$^\dagger$ [ $\msun\,yr^{-1}$] &&&& \\
$> $10  &  58 & 4.47 $\pm$  1.54 &   9.29 $\pm$  4.42 &   {\bf 5.54 $\pm$  1.67} &   2.08 $\pm$  1.32   & 19 &    1.97 $\pm$  0.86 &   0.70 $\pm$  0.34        & 31 &   1.96 $\pm$  1.02  \\
1 -- 10   &  153 & {\bf 1.48 $\pm$  0.19} &   {\bf 3.20 $\pm$  0.63} &   {\bf 3.07 $\pm$  0.49} &   {\bf 2.15 $\pm$  0.40 } & 31&    4.82 $\pm$  2.55 &   1.15 $\pm$  0.73        & 40 &  6.47 $\pm$  4.93  \\
\hline
\hline
\label{tab:ratio}
\end{tabular}
\end{center} 
Notes: 
Given the bias against individual under-detection of multiple lines,
and the mass-independent treatment of \nii\ correction and the lack of dust-extinction correction,
the absolute values reported in this table
should be used with caution. 
They are listed to show the trends of the various line ratios between different
subsamples. 
Here N${_i}$ refer to the numbers of pairs used in each stack,
which correspond to the redshift range where the relevant lines are covered:
0.69 $< z_1 < $ 1.51 (\hbeta, \oiii, \halpha,\sii), 1.28 $< z_2 < $ 1.45 (\oii, \halpha, \sii), and 1.28 $ < z_3 < $ 2.29 (\oii, \oiii). 
Only spectra with both G102 and G141 coverages are included in the stack.
Line ratios with S/N $>$ 3 are highlighted in boldface.  
*: Since the grism spectral resolution is not sufficient for \halpha\ and [NII] separation, 
here \halpha\ refers to the manually corrected \halpha\ flux,
with an [NII] correction of 15\% applied to the [NII]$+$\halpha\ flux. 
This value is the mean correction value adopted from \citep{dominguez13, masters14}. \\
$\dagger$: No pair with SFR $<$ 1 $\msun\,yr^{-1}$ falls in the selected redshift ranges listed above. 
\end{table*}

\begin{table*}
\begin{center}
\caption{Metallicity based on the stacked spectra}
\begin{tabular}{ccc}
\hline
\hline
galaxy type  & N${_3}$ & 12$+$log(O/H) \\
& & \citet{kk04} \\
\hline
\hline
control sample       & 758   & 8.82 $\pm$  0.03 \\
\hline
         all pairs          & 87  &     8.59$^{+0.57}_{-0.32}$ \\
             disturbed  & 26  &  8.81$^{+0.17}_{-0.12}$  \\
  merging  & 8&  8.86$^{+0.12}_{-0.10}$ \\
   secure   & 24  &   8.78$^{+0.45}_{-0.18}$  \\
    wide   & 55  &   8.67$^{+0.60}_{-0.34}$ \\
\hline
EW [\AA] && \\
   $> 500$     & 11 & 7.48$^{+0.19}_{-0.17}$ \\
   100 -- 500  & 39 &  8.92$^{+0.15}_{-0.13}$    \\
   $< $100     & 17 &  8.71$^{+0.26}_{-0.21}$\\
\hline
SFR [ $\msun\,yr^{-1}$] && \\
$> $10   & 31 &   9.05$^{+0.02}_{-0.21}$ \\
1 -- 10   & 40  &  8.74$^{+0.47}_{-0.27}$\\
\hline
\hline
\label{tab:met}
\end{tabular}
\end{center} 
Notes: Values derived from the stacked spectra in redshift range, 1.28 $ < z_3 < $ 2.29, 
where\oii, \oiii, \hbeta\ lines are covered. 
\end{table*}

\begin{table*}
\begin{center}
\caption{Emission Line Galaxies Pair fraction}
\begin{tabular}{lccccccccc}
\hline
\hline
\vspace{3pt} redshift range &  $N_{\rm pairs}$ & $<z>$ & f$_{\rm total}$ (\%) & $N_{\rm major}$ &$<z>$  & $f_{\rm major}$ (\%) & $N_{\rm minor}$ &$<z>$  & $f_{\rm minor}$ (\%) \\
\hline
\vspace{2pt} 0.28 $< z \leq $ 0.60  & 73  & 0.49 & 9.4$^{+0.8}_{-0.7}$    & 40   &0.48& 4.0$^{+0.6}_{-0.5}$   & 33 &0.46&2.9$^{+0.6}_{-0.4}$   \\ 
\vspace{2pt} 0.60 $< z \leq $ 0.90  &  82 & 0.73  &  9.4$^{+0.7}_{-0.6}$   & 58   &0.74& 6.4$^{+0.6}_{-0.5}$      &24 &0.69&3.0$^{+0.4}_{-0.3}$  \\
\vspace{2pt} 0.90 $< z \leq $ 1.20  & 163& 1.07 &  12.8$^{+0.6}_{-0.6}$   & 108  &1.07& 8.3$^{+0.5}_{-0.4}$       &55 &1.08&4.5$^{+0.4}_{-0.3}$ \\
\vspace{2pt} 1.20 $< z \leq $ 1.50  & 172 & 1.34 &  13.7$^{+0.7}_{-0.6}$  & 119 &1.35& 9.1$^{+0.6}_{-0.5}$       &53 &1.34&4.6$^{+0.4}_{-0.3}$ \\
\vspace{2pt} 1.50 $< z \leq $ 1.60$^*$  & 15 & 1.53  &  7.0$^{+1.3}_{-0.9}$    & 8   & 1.52& 3.9$^{+1.0}_{-0.6}$       &7  &1.52&3.1$^{+1.0}_{-0.6}$  \\
\hline
\hline
\label{tab:frac}
\end{tabular}
\end{center} 
Notes: $<z>$ refers to the average $z$ in the relevant bins.
The fraction shown here has been completeness-corrected following the recipes described in Sec~\ref{sec:frac}.
The major and minor pairs are selected by their H-band flux ratios (See \S\ref{sec:sample}).
*: $z=$1.6 is the redshift limit below which the completeness correction, mainly based by \halpha$_{\rm, raw}$, is more reliable. 
Fitting the data up to $z= 1.6$ yields a linear correlation of 
$f=  (0.08\pm\,0.01)\times(1+ z)^{(0.58\pm\,0.17)}$ for the whole pair sample,
$f = (0.04\pm\,0.01) \times(1+z)^{(0.77\pm\,0.22)}$ for the major pairs,
and $f = (0.03\pm\,0.01) \times(1+z)^{(0.35\pm\,0.30)}$ for the minor pairs. 
\end{table*}

\end{CJK}
\end{document}